\documentclass[pre,aps,twocolumn,superscriptaddress,showpacs]{revtex4}

\usepackage[dvips]{graphicx}
\usepackage{amssymb,amsfonts,amsmath}
\usepackage{color}
\usepackage{ulem}
\usepackage{hyperref}

\newcommand{\Tr}{{\rm Tr}\,}
\newcommand{\intg}{\mkern-3mu\int\mkern-6mu}
\newcommand{\JJ}{\underline{\underline{\mathbf{J}}}_{\mkern2mu 2}}

\begin{document}

\title{Dynamic Pair Correlations and Superadiabatic Forces\\in a Dense Brownian Liquid}

\author{Thomas Schindler}
\affiliation{Theoretische Physik II, Physikalisches Institut, 
  Universit{\"a}t Bayreuth, D-95440 Bayreuth, Germany}

\author{Matthias Schmidt}
\affiliation{Theoretische Physik II, Physikalisches Institut, 
  Universit{\"a}t Bayreuth, D-95440 Bayreuth, Germany}

\date{\today}

\begin{abstract}
We study dynamic two-body correlation functions, i.e.\ the two-body density,
the current-density correlator or van Hove current, and the current-current correlator
in Brownian dynamics computer simulations of a dense Lennard-Jones
bulk liquid. The dynamic decay of the correlation shells of the two-body
density is examined in detail. Inner correlation shells decay faster
than outer correlation shells, whereas outer correlation shells remain
stable for increasing times. Within a dynamic test particle picture the mechanism is assumed to be triggered
by the dislocation of the self particle, which releases the confinement
of the surrounding correlation shells. We present a division of the
van Hove current into an adiabatic and a superadiabatic contribution. The magnitude of the adiabatic
van Hove current is found to exceed that of the total van Hove current,
which is consistent with dynamic density functional
theory overestimating the speed of the dynamics.
The direction of the superadiabatic van Hove current opposes that of the total van Hove current.
The current-current correlator reveals detailed insight in the collisions of the particles.
We find a large static nearest-neighbor peak, which results from colliding particles and different dynamic peaks, that are attributed to consecutive collisions.
\end{abstract}

\pacs{61.20.Gy, 61.20.Ja, 05.20.Jj}

\maketitle

\section{Introduction}
A very powerful tool to determine static equilibrium properties of a many-body system
is classical density functional theory (DFT) \citep{DFT-Mermin,DFT-Evans79}.
For a comprehensive introduction to the field and a gamut of current developments we refer the reader to Ref.\ \citep{DFT_overview_16}.
In DFT the equilibrium one-body density distribution of a system
with arbitrary external one-body potential is obtained by minimizing the grand
potential functional \citep{DFT-Evans79}. One major advantage of
DFT is the fact that the grand potential can be split into one part
specific for the interparticle interactions and an additive contribution
that only depends on the external field. This implies that -- once a
suitable approximation for the intrinsic free energy functional is found for a
specific model -- the theory can be applied to any external field in order to
obtain equilibrium properties. One extension of DFT to non-equilibrium
systems is dynamic density functional theory (DDFT) \citep{DDFT99,Archer-Evans04}.
Here the equal-time two-body density, which e.g.\ describes the structure
of the liquid, of a system out-of-equilibrium is assumed to be equal to a corresponding equilibrium version.
Hereby it is implied that the relaxation of density correlations is faster than the dynamics of the density profile itself.
However the validity of this approximation is not clear but it provides a practical solution that allows one
to use the equilibrium grand potential functional to obtain the dynamics of the density profile in non-equilibrium situations.
The method produces qualitatively good results compared to Brownian dynamics (BD) simulations
but tends to overestimate the rate of relaxation processes in a variety of situations \citep{DDFT99,DDFT-HRods,Fortini-et-al,dTPL,dTPL2}.
When applying the dynamic test particle limit (TPL) \citep{dTPL,dTPL2}
the theory can be used to examine the well-known van Hove function \citep{GvH} and bulk dynamics.
Here the overestimated dynamics manifest for instance in a wrong description of long-time diffusion.
Stopper $et\,al.$ recently addressed the problem with an empirically biased version of DDFT \citep{Stopper_Marolt_15,Stopper_Roth_15}.

Another extension of DFT, which aims at an exact description of non-equilibrium systems, is the recently presented power functional
theory (PFT) \citep{PFT13}. In PFT the one-body particle current is treated as the fundamental
variable rather than the density distribution. The current is calculated
from a minimization condition of a free power functional.
The development of concrete approximations of the power functional for specific models is an ongoing task \citep{PFT13,NOZ,active}.
PFT is exact in the sense that there is no adiabatic assumption and the
full non-equilibrium behavior of the system is captured. The difference
between the adiabatic DDFT current and the full non-equilibrium current
obtained via PFT is defined as the superadiabatic current.
When used in the context of the dynamic TPL in the PFT framework \citep{PFT-DTPL15},
the superadiabatic current is related to direct time correlation
functions via a non-equilibrium Ornstein-Zernike equation \citep{NOZ}.
These direct correlation functions contain memory of the past motion
of the system and can be represented as functional derivatives
of the excess (over ideal) free power functional with respect to density
and/or current \citep{PFT14}. Hence, knowledge of the correlation functions
and memory functions is crucial to obtaining future explicit approximations
to power functionals and making PFT a readily applicable tool.  

In the present work we investigate the dynamic correlations in an equilibrium
bulk Lennard-Jones (LJ) system via BD simulations.
By examining the bulk system we intend to obtain the inherent features of the correlation functions,
that we expect to recover in any inhomogeneous and non-equilibrium system as well.
The choice for this particular model was made as it is a prime example of a simple fluid.
The LJ fluid has been widely studied theoretically and is naturally applied to describe atomic fluids such as noble gases.
Thus, the properties and phase behavior of the model are well-known (see e.g.\ \citep{LJ-PD-rc4.4,LJ}).
Between the triple point temperature and the critical temperature the phase diagram is separated by the binodal into
a low density gas phase and a liquid phase with intermediate density.
We examine a liquid state point close to the triple point in order to potentially maximize the influence of the interparticle forces.

The arguably simplest among the dynamic correlation functions is the two-body density, which is
(up to a multiplicative constant) equivalent to the van Hove function and, in the equal time
case, to the radial distribution function (RDF). In colloidal systems
the intermediate scattering function, which is the spatial Fourier
transform of the van Hove function, can be measured in neutron scattering
experiments \citep{hansen_mcdonald} or with video microscopy (see
e.g.\ \citep{video-microscopy-Valentine,video-microscopy-Murray}).
This makes the van Hove function a useful tool for comparing theory and experiments
and for analyzing complex phenomena, such as formation of transient networks \citep{transient}.

The fundamental theory for the long-range decay of the $static$ two-body density \citep{decay93,Asymptotic1}
predicts monotonic exponential decay at low densities and temperatures and oscillatory exponential decay at high densities and temperatures.
The two regions in the phase diagram are separated by the so-called Fisher-Widom line \citep{FW}.
Very recent experimental and simulation work has further validated the theory for asymptotic decay
of correlations for colloidal systems \citep{Statt_Pinchaipat_16}.
Furthermore, the universality of the decay was shown in Ref.\ \citep{Klapp_Zeng_08}.
Although the decay scenario is well-investigated in the static case,
it appears that it has not been extended to the behavior of dynamic correlations so far.

The second investigated correlation function is the van Hove current, which is a density-current
correlator and particularly interesting in the frameworks of DDFT and PFT \citep{dTPL,dTPL2,PFT-DTPL15}.
The connection to the van Hove function is via a continuity
equation, which can therefore be used to obtain the time evolution of
the van Hove function. The van Hove current can be identified with a one-body
current via the dynamic TPL \citep{dTPL}. Thus, it acts as
the fundamental variable in the PFT framework applied to bulk fluids
\citep{PFT-DTPL15}. A further objective of the present work is to
obtain the adiabatic contribution to the van Hove current and to compare
it to the total van Hove current. Thereby we aim at determining the
accuracy of the adiabatic assumption for the examined model and state point.
The third correlation function -- the current-current correlator -- is in general a tensorial
quantity that relates the velocities of two particles. We present
a straightforward implementation to calculate the non-zero components
in bulk. The Fourier transform of the longitudinal current-current
correlator is the velocity autocorrelation function \citep{hansen_mcdonald}.
Recently this function has been measured via the time derivative of
the intermediate scattering function \citep{current-corr-exp}.

The paper is organized as follows. In Sec.\ \ref{sec:Theory} we
present the system and the simulation details (Sec.\ \ref{sub:Brownian-Dynamics})
followed by an introduction of the correlation functions (Sec.\ \ref{sub:2-body-corr-fkts}).
In bulk several symmetries apply, which we exploit
to simplify the examination of the correlation functions (Sec.\ \ref{sub:Symm-and-Cosy}).
We derive continuity relations between the correlation functions (Sec.\ \ref{sub:CEq-theory}),
which are used to link the behavior of the different correlation functions
and to check the consistency of our numerical data. A method to calculate
the adiabatic and superadiabatic van Hove current is presented, which combines
the dynamic TPL (Sec.\ \ref{sub:TPL}) and the explicit construction
of the adiabatic state (Sec.\ \ref{sub:Superadiabatic-Force}).
In order to gain deeper insight into the decay mechanisms of the correlation functions
we study a reference system without particle-particle interactions,
where particles move via ideal diffusive motion (Sec.\ \ref{sub:Relaxing-System}).
The asymptotic expansion \citep{decay93,Asymptotic1} for large distances of the two-body density is briefly reviewed in Sec.\ \ref{sub:log-rho};
this has so far been applied only to the static two-body density.
Section \ref{sec:Results} presents the numerical results for the correlation functions.
We show that the self parts of the correlation functions in the examined LJ liquid exhibit the same features as in the ideal gas (Sec.\ \ref{sub:Results-Self}).
Furthermore we identify the mechanisms determining the qualitative shapes of the distinct correlation functions
and the relations between those (Sec.\ \ref{sub:Results-Distinct}).
The results of the adiabatic construction that yield the adiabatic and
superadiabatic van Hove currents are presented in Sec.\ \ref{sub:Results-Jad}.
Concluding remarks and suggestions for further investigations are given in Sec\ \ref{sec:Conclusion}.
In Sec.\ \ref{sec:Appendix} we present supplementary calculations of divergences of isotropic vector
fields and diagonal tensor fields in spherical coordinates (Sec.\ \ref{sub:Del*JvH}).
We verify the consistency of our numerical
data for the correlation functions by means of the continuity equations (Sec.\ \ref{sub:CEq-numerical})
and the analytic calculation of the self correlation functions for
the ideal diffusive motion is performed (Sec.\ \ref{sub:J-self-re}).
These technical results constitute a resource for carrying out actual computations.

\section{Theory and methods\label{sec:Theory}}
\subsection{Brownian dynamics and Lennard-Jones model\label{sub:Brownian-Dynamics}}

We simulate a bulk fluid of LJ particles with particle density
$\rho$ in an implicit viscous solvent at temperature $T$. Simulations
are carried out with $N$ particles in a cubic box of volume $V=N/\rho$.
Periodic boundary conditions are applied in all three spatial directions
in order to minimize finite size effects. Numerical results for the correlation
functions of interest (cf.\ Sec.\ \ref{sub:2-body-corr-fkts})
are obtained as time averages of single long runs.

The trajectories $\mathbf{r}_{i}(t)$, i.e.{\ }the
position of each particle $i=1,...,N$ at time $t$, are obtained in discrete
time steps $\Delta t$, where
\begin{equation}
\mathbf{r}_{i}(t+\Delta t)=\mathbf{r}_{i}(t)+\dot{\mathbf{r}}_{i}(t)\Delta t,
\label{eq:trajectory-1}
\end{equation}
with $\dot{\mathbf{r}}_{i}(t)$ being the displacement velocity between the times $t$ and $t+\Delta t$.
We consider the overdamped Stokes limit, which implies that inertia
of the particles is neglected. Thus, the displacement velocity $\dot{\mathbf{r}}_{i}(t)$
of each particle is directly proportional to the force acting on that particle,
\begin{equation}
\xi\dot{\mathbf{r}}_{i}(t)=\mathbf{F}_{i}(t)=\sum_{j\neq i}^{N}\mathbf{F}_{ij}(t)+\mathbf{R}_{i}(t),\label{eq:overdamp}
\end{equation}
where $\xi$ is the friction coefficient and the sum is over all
other particles $j\neq i$. The total force $\mathbf{F}_{i}(t)$ acting
on particle $i$ consists of pairwise intrinsic contributions $\mathbf{F}_{ij}(t)$
of particle $i$ with each other particle $j$ and a random force
$\mathbf{R}_{i}(t)$ modeling the interaction with the solvent.

The force $\mathbf{F}_{ij}(t)$, that particle $j$ exerts on particle $i$, is the negative
gradient of the 6-12-LJ pair potential with respect to $\mathbf{r}_{i}(t)$,
\begin{equation}
\mathbf{F}_{ij}(t)=\begin{cases}
\epsilon\left[48\left(\frac{\sigma}{r_{ij}}\right)^{12}-24\left(\frac{\sigma}{r_{ij}}\right)^{6}\right]\frac{\mathbf{r}_{ij}}{r_{ij}^{2}} & \,\,\,\,\,r_{ij}<r_{c}\\
0 & \,\,\,\,\,\mathrm{otherwise,}
\end{cases}
\end{equation}
where $r_{ij}\equiv\left|\mathbf{r}_{ij}\right|\equiv\left|\mathbf{r}_{i}(t)-\mathbf{r}_{j}(t)\right|$
is the distance between the particles $i$ and $j$ at time $t$. In
order to minimize computational cost, we truncated and shifted the pair
potential at $r_{c}=4\sigma$, following Ref.\ \citep{frencel_smit}.

The effect of collisions of particle $i$ with (implicit) solvent
molecules is modeled as a random force $\mathbf{R}_{i}(t)$. The
collisions are assumed to be uncorrelated with each other at different
times. Furthermore, the forces that act on different particles are
uncorrelated with each other. The strength of the random force is
determined via its autocorrelation function,
\begin{equation}
\left\langle \mathbf{R}_{i}(t)\mathbf{R}_{j}(t')\right\rangle =2k_{\mathrm{B}}T\xi\delta_{ij}\delta(t-t')\underline{\underline{\mathbf{I}}}.\label{eq:corr-force}
\end{equation}
Here the angles denote an average over realizations of the
random forces, the product of vectors (without dot) denotes the dyadic
product yielding the $3\times3$ unit matrix $\underline{\underline{\mathbf{I}}}$,
$k_{\mathrm{B}}$ is Boltzmann's constant, $\delta(\cdot)$ is the
Dirac delta distribution, and $\delta_{ij}$ is the Kronecker delta
symbol. In order to attain these properties the random force is chosen to
be a random vector with Gaussian probability distribution $p(\mathbf{R}_{i}(t))$ \citep{Gauss},
where each component has a variance of $\sigma_{R}^{2}=2\xi k_{\mathrm{B}}T/\Delta t$,
\begin{equation}
p(\mathbf{R}_{i}(t))=(2\pi\sigma_{R}^{2})^{-3/2}\exp\left(-\frac{\mathbf{R}_{i}^{2}(t)}{2\sigma_{R}^{2}}\right).\label{eq:ran_Force-1}
\end{equation}
The probability distribution is normalized to unity as its integral
over all random forces is a Gaussian integral, $\int\mathrm{d}\mathbf{R}_{i}(t)p(\mathbf{R}_{i}(t))=1$.
Note that the factor $1/\Delta t$ acts as a discretized version of $\delta(t-t')$ in Eq.~(\ref{eq:corr-force}).

We use the particle diameter $\sigma$ as the unit of length, the
energy constant $\epsilon$ as the unit of energy, and the friction
constant $\xi$ as the unit of friction.
This implies a time scale of $\tau_{0}=\xi\sigma^{2}/\epsilon$ which is used as the unit of time.
Note that a corresponding Brownian time scale is $\tau_{\mathrm{B}}=\epsilon \tau_{0}/(k_{\mathrm{B}}T)$.
One-body density and current are then measured in units of $\sigma^{-3}$ and $\tau_{0}^{-1}\sigma^{-2}$, respectively.
The two-body density, the van Hove current, and the current-current correlator
are measured in units of  $\sigma^{-6}$, $\tau_{0}^{-1}\sigma^{-5}$, and $\tau_{0}^{-2}\sigma^{-4}$, respectively.
For convenience we set $\sigma \equiv 1$, $\epsilon \equiv 1$, and $\tau_{0}\equiv 1$ when presenting our numerical data in the results section below.
Simulations are carried out with $N=500$ particles at temperature $T=0.8\epsilon/k_{\mathrm{B}}$
and number density $\rho=0.84/\sigma^3$. This state point is a fluid state
close to the coexisting liquid at the LJ triple point $T_{\mathrm{tr}}=0.7\epsilon/k_{\mathrm{B}}$
and $\rho_{\mathrm{tr,l}}=0.84/\sigma^3$ \citep{LJ-TrP}.
Leote de Carvalho $et\,al.$ have examined the liquid structure of a supercritical LJ fluid with $r_c=2.5$
at the state point $T=1.2\epsilon/k_{\mathrm{B}}$ and $\rho=0.715/\sigma^3$ \citep{Leote-de-Carvalho-Evans94}.
We choose the time step $\Delta t=5\cdot10^{-5}\tau_{0}$.

\subsection{Dynamic two-body correlation functions\label{sub:2-body-corr-fkts}}  

We define the one-body density operator as
\begin{equation}
\hat{\rho}(\mathbf{r},t)\equiv\sum_{i=1}^{N}\delta(\mathbf{r}-\mathbf{r}_{i}(t)),\label{eq:rhoop}
\end{equation}
which describes the (unnormalized) probability density to find a particle at position $\mathbf{r}$
at time $t$ in microstate $(\mathbf{r}_{1}(t),...,\mathbf{r}_{N}(t))$. The operator of the associated particle current is
\begin{equation}
\hat{\mathbf{J}}(\mathbf{r},t)\equiv\sum_{i=1}^{N}\mathbf{v}_{i}(t)\delta(\mathbf{r}-\mathbf{r}_{i}(t)),\label{eq:Jop}
\end{equation}
describing the (vectorial) current of particles at position $\mathbf{r}$
at time $t$. Here we use the velocity $\mathbf{v}_{i}(t)$ as the
symmetric derivative of the position of particle $i$ with respect
to time \citep{Fortini-et-al,Bernreuther_Schmidt},
\begin{equation}
\mathbf{v}_{i}(t)\equiv\frac{\mathbf{r}_{i}(t+\Delta t)-\mathbf{r}_{i}(t-\Delta t)}{2\Delta t}.
\end{equation}
The symmetrical velocity is related to the displacement velocity $\dot{\mathbf{r}}_{i}(t)$,
defined in Eq.{\ }(\ref{eq:trajectory-1}),
via
\begin{equation}
\mathbf{v}_{i}(t)=\frac{\dot{\mathbf{r}}_{i}(t)+\dot{\mathbf{r}}_{i}(t-\Delta t)}{2}.\label{eq:v-rdot}
\end{equation}
For brevity we introduce the shorthand notation $(1)\equiv(\mathbf{r}_1,t_1)$ and $(2)\equiv(\mathbf{r}_2,t_2)$ for space-time points.
Here $\mathbf{r}_{1}$ and $\mathbf{r}_{2}$ (without time
arguments) denote two fixed points in the system rather than the positions
of particles $\mathbf{r}_{i}(t)$ with $i=1,2$.
Two-body correlation functions can now be formed by multiplying
two of the one-body operators (\ref{eq:rhoop}) and (\ref{eq:Jop}) above and averaging.

The (scalar) two-body density $\rho_{2}$ is formed by two density operators \citep{hansen_mcdonald},
\begin{align}
\rho_{2}(1,2)&\equiv\bigl\langle\hat{\rho}(1)\hat{\rho}(2)\bigr\rangle\label{eq:rho2}\\
&=\left\langle \sum_{i=1}^{N}\sum_{j=1}^{N}\delta(\mathbf{r}_{1}-\mathbf{r}_{i}(t_{1}))\delta(\mathbf{r}_{2}-\mathbf{r}_{j}(t_{2}))\right\rangle ,\nonumber
\end{align}
where the angles denote an average over initial conditions
and realizations of random forces. $\rho_{2}$ measures the joint
probability to find a particle at position $\mathbf{r}_{1}$ at time
$t_{1}$ and a particle at position $\mathbf{r}_{2}$ at time $t_{2}$.
The two-body density is proportional to the van Hove function $G_{\mathrm{vH}}(1,2)$
\citep{GvH}, which can be viewed as the local and time resolved particle
density at space-time point 1, given that there is a particle located at space-time point 2.

Following the notation of Ref.\ \citep{PFT14} we denote the averaged
product of a density operator and a current operator as van Hove current
$\mathbf{J}_{\mathrm{vH}}$. We use the convention that $t_{2}$
denotes the earlier point in time, i.e.{\ }$t_{2}\leq t_{1}$.
Therefore, we can define two different correlation functions -- a
backward van Hove current $\mathbf{J}_{\mathrm{vH}}^{\mathrm{back}}$
and a forward van Hove current $\mathbf{J}_{\mathrm{vH}}^{\mathrm{for}}$,
via
\begin{align}
& \mathbf{J}_{\mathrm{vH}}^{\mathrm{back}}(1,2)\equiv\bigl\langle\hat{\rho}(1)\hat{\mathbf{J}}(2)\bigr\rangle\label{eq:Jback}\\
& =\left\langle \sum_{i=1}^{N}\sum_{j=1}^{N}\delta(\mathbf{r}_{1}-\mathbf{r}_{i}(t_{1}))\mathbf{v}_{j}(t_{2})\delta(\mathbf{r}_{2}-\mathbf{r}_{j}(t_{2}))\right\rangle ,\nonumber 
\end{align}
\begin{align}
& \mathbf{J}_{\mathrm{vH}}^{\mathrm{for}}(1,2)\equiv\bigl\langle\hat{\mathbf{J}}(1)\hat{\rho}(2)\bigr\rangle\label{eq:Jfor}\\
& =\left\langle \sum_{i=1}^{N}\sum_{j=1}^{N}\mathbf{v}_{i}(t_{1})\delta(\mathbf{r}_{1}-\mathbf{r}_{i}(t_{1}))\delta(\mathbf{r}_{2}-\mathbf{r}_{j}(t_{2}))\right\rangle .\nonumber 
\end{align}

The (tensorial) current-current correlator $\JJ$ is formed by the average of a dyadic product of two current operators
(see e.g.~\citep{PFT14} for the present case of BD and \citep{hansen_mcdonald} for molecular dynamics):
\begin{align}
& \JJ(1,2)\equiv\bigl\langle\hat{\mathbf{J}}(1)\hat{\mathbf{J}}(2)\bigr\rangle\label{eq:J2}\\
 & =\left\langle \sum_{i=1}^{N}\sum_{j=1}^{N}\mathbf{v}_{i}(t_{1})\delta(\mathbf{r}_{1}-\mathbf{r}_{i}(t_{1}))\mathbf{v}_{j}(t_{2})\delta(\mathbf{r}_{2}-\mathbf{r}_{j}(t_{2}))\right\rangle .\nonumber 
\end{align}
The trace of this tensor reads
\begin{align}
& \Tr\JJ(1,2)=\bigl\langle\hat{\mathbf{J}}(1)\cdot\hat{\mathbf{J}}(2)\bigr\rangle\label{eq:Trace}\\
& =\left\langle \sum_{i=1}^{N}\sum_{j=1}^{N}\mathbf{v}_{i}(t_{1})\delta(\mathbf{r}_{1}-\mathbf{r}_{i}(t_{1}))\cdot\mathbf{v}_{j}(t_{2})\delta(\mathbf{r}_{2}-\mathbf{r}_{j}(t_{2}))\right\rangle .\nonumber 
\end{align}

Each of the four correlation functions (\ref{eq:rho2}) -- (\ref{eq:J2})
and the trace of the current-current correlator (\ref{eq:Trace}) can be separated into
a self and a distinct part by dividing the double sum into two sums
-- one sum with terms where $i=j$ and one sum with terms where $i\neq j$.
Taking the two-body density as an example, this reads 
\begin{equation}
\rho_{2}(1,2)\equiv\rho_{2}^{\mathrm{self}}(1,2)+\rho_{2}^{\mathrm{dist}}(1,2),\label{eq:separation-self-dist}
\end{equation}
where
\begin{equation}
\mkern-11mu\rho_{2}^{\mathrm{self}}(1,2)=\left\langle \sum_{i=1}^{N}\delta(\mathbf{r}_{1}-\mathbf{r}_{i}(t_{1}))\delta(\mathbf{r}_{2}-\mathbf{r}_{i}(t_{2}))\right\rangle ,\label{eq:rho2_self}
\end{equation}
\begin{equation}
\mkern-11mu\rho_{2}^{\mathrm{dist}}(1,2)=\left\langle \sum_{i=1}^{N}\sum_{j\neq i}^{N}\delta(\mathbf{r}_{1}-\mathbf{r}_{i}(t_{1}))\delta(\mathbf{r}_{2}-\mathbf{r}_{j}(t_{2}))\right\rangle .\label{eq:rho2_dist}
\end{equation}
Analogous splittings hold for the van Hove currents $\mathbf{J}_{\mathrm{vH}}^{\mathrm{back}}$,
$\mathbf{J}_{\mathrm{vH}}^{\mathrm{for}}$, the current-current correlator
$\JJ$, and its trace $\Tr\JJ$.
Henceforth we indicate self and distinct parts of functions with a
superscript $\alpha\equiv\mathrm{self},\mathrm{dist}$. The splitting
into self and distinct parts allows to discriminate between the behavior
of a single (tagged) particle in the presence of its surrounding particles
(self part) and the behavior of the surrounding particles in the neighborhood
of the tagged particle (distinct part). 

For the two-body density the normalization condition
\begin{equation}
\intg\mathrm{d}\mathbf{r}_{1}\intg\mathrm{d}\mathbf{r}_{2}\rho_{2}(1,2)=\left\langle \intg\mathrm{d}\mathbf{r}_{1}\hat{\rho}(1)\intg\mathrm{d}\mathbf{r}_{2}\hat{\rho}(2)\right\rangle =N^{2}\label{eq:norm-rho2}
\end{equation}
holds, where the integrals are performed over the system (i.e.\ the
simulation box) volume. For the second equality in Eq.\ (\ref{eq:norm-rho2})
we employ the normalization condition for the one-body density operator,
$\int\mkern-4mu\mathrm{d}\mathbf{r}\hat{\rho}(\mathbf{r},t)=N$.
For the self and distinct parts we obtain in analogy to Eq.{\ }(\ref{eq:norm-rho2})
\begin{equation}
\intg\mathrm{d}\mathbf{r}_{1}\intg\mathrm{d}\mathbf{r}_{2}\rho_{2}^{\mathrm{self}}(1,2)=N,\label{eq:norm-self}
\end{equation}
\begin{equation}
\intg\mathrm{d}\mathbf{r}_{1}\intg\mathrm{d}\mathbf{r}_{2}\rho_{2}^{\mathrm{dist}}(1,2)=N(N-1).\label{eq:norm-dist}
\end{equation}

\subsection{Symmetries and choice of coordinate system\label{sub:Symm-and-Cosy}}

In case of an equilibrium bulk fluid, temporal translational, spatial trans\-lational, and rotational symmetries apply.
The translational symmetries imply that the correlation functions of Sec.\ \ref{sub:2-body-corr-fkts}
will not depend explicitly on positions $\mathbf{r}_{1}$ and $\mathbf{r}_{2}$
and points in time $t_{1}$ and $t_{2}$, but only on the difference
vector $\mathbf{r}\equiv\mathbf{r}_{1}-\mathbf{r}_{2}$ and time difference
$\tau\equiv t_{1}-t_{2}$. The partial derivative with respect to
$\tau$ is related to the partial derivatives with respect to $t_{1}$
and $t_{2}$ via
\begin{equation}
\partial_{\tau}=\partial_{t_{1}}=-\partial_{t_{2}}.\label{eq:dt}
\end{equation}
The gradient with respect to $\mathbf{r}$ is related to the gradients
with respect to $\mathbf{r}_{1}$ and $\mathbf{r}_{2}$ via 
\begin{equation}
\nabla=\nabla_{1}=-\nabla_{2}.\label{eq:dr}
\end{equation}
Furthermore the two-body density depends only on the distance $r\equiv\left|\mathbf{r}\right|$
but not on the orientation of $\mathbf{r}$ because of the rotational
symmetry. Hence, we sample $\rho_{2}^{\alpha}$ as a function of $r$
and $\tau$.

The bulk van Hove current is an isotropic vector field, which implies
that $\mathbf{J}_{\mathrm{vH}}^{\alpha}(\mathbf{r},\tau)$ is parallel (or antiparallel)
to $\mathbf{r}$ and its magnitude only depends on $r$ and $\tau$.
In order to exploit this symmetry we use spherical coordinates with radial
unit vector $\mathbf{e}_{r}\equiv\mathbf{r}/r$, azimuthal unit vector
$\mathbf{e}_{\varphi}\equiv\mathbf{e}_{r}\times\mathbf{e}_{z}/\left|\mathbf{e}_{r}\times\mathbf{e}_{z}\right|$ (where $\mathbf{e}_{z}$ is the unit vector $z$-direction of the simulation box),
and polar unit vector $\mathbf{e}_{\theta}\equiv\mathbf{e}_{r}\times\mathbf{e}_{\varphi}$.
(The case $\mathbf{e}_{r}\parallel\mathbf{e}_{z}$ is sufficiently
unlikely.) In this coordinate system the van Hove current reads
\begin{equation}
\mathbf{J}_{\mathrm{vH}}^{\alpha}(\mathbf{r},\tau)=J_{r}^{\alpha}(r,\tau)\mathbf{e}_{r}.\label{eq:JvH-radial}
\end{equation}
This equation holds both for the forward and for the backward van
Hove current. Due to the symmetry arguments given above, the transversal
components $J_{\varphi}^{\alpha}$ and $J_{\theta}^{\alpha}$ vanish.
Nevertheless we have sampled these components as a consistency check
and found them to be zero within the statistical fluctuations.

The bulk current-current correlator is an isotropic tensor with only
radial-radial non-zero component $J_{rr}^{\alpha}(r,\tau)\equiv\mathbf{e}_{r}\mathbf{e}_{r}\mathbf{:}\JJ^{\alpha}(\mathbf{r},\tau)$
and two equal trans\-versal-trans\-versal non-zero components $J_{tt}^{\alpha}(r,\tau)\equiv\mathbf{e}_{\varphi}\mathbf{e}_{\varphi}\mathbf{:}\JJ^{\alpha}(\mathbf{r},\tau)=\mathbf{e}_{\theta}\mathbf{e}_{\theta}\mathbf{:}\JJ^{\alpha}(\mathbf{r},\tau)$
\citep{hansen_mcdonald}; here the colon indicates a double tensor contraction.
Hence, in the spherical coordinate system $\JJ^{\alpha}$ reads
\begin{equation}
\JJ^{\alpha}(\mathbf{r},\tau)=J_{rr}^{\alpha}(r,\tau)\mathbf{e}_{r}\mathbf{e}_{r}+J_{tt}^{\alpha}(r,\tau)(\mathbf{e}_{\varphi}\mathbf{e}_{\varphi}+\mathbf{e}_{\theta}\mathbf{e}_{\theta}).\label{eq:J2-isotropic}
\end{equation}
We checked the validity of the symmetry arguments by sampling
each of the nine components $J_{kl}^{\alpha}$, with $k,l=r,\varphi,\theta$, individually.
The off-diagonal components (with $k\neq l$) were found to be zero
and the two transversal-transversal components $J_{\varphi\varphi}$
and $J_{\theta\theta}$ were found to be equal, yielding a current-current
correlator of the form of Eq.\ (\ref{eq:J2-isotropic}). The
results presented below for the transversal-transversal component (Sec.\ \ref{sec:Results})
are the arithmetic mean of the azimuthal-azimuthal component and the
polar-polar component $J_{tt}^{\alpha}=(J_{\varphi\varphi}^{\alpha}+J_{\theta\theta}^{\alpha})/2$
to reduce noise of the data.

\begin{figure}
\begin{centering}
\includegraphics[width=8.0cm,angle=0]{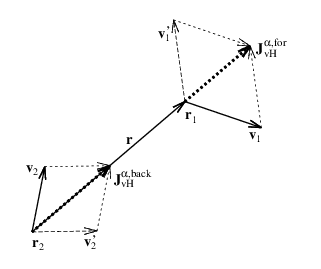}
\end{centering}
\caption{\label{fig:sketch-cosy}Illustration for the canceling of transversal contributions to the
van Hove current in isotropic systems. Symbols are defined in the text.}
\end{figure}

For an illustration of the symmetries of the van Hove current,
consider a configuration with two particles (or one and the same particle)
at space-time points $\mathbf{r}_{1},t_{1}$ and $\mathbf{r}_{2},t_{2}$
with velocities $\mathbf{v}_{1}$ and $\mathbf{v}_{2}$ as sketched in Fig.~\ref{fig:sketch-cosy}.
Consider also a further configuration with two particles at the same space-time
points but with velocities $\mathbf{v}_{1}'$ and $\mathbf{v}_{2}'$,
which are the velocities $\mathbf{v}_{1}$ and $\mathbf{v}_{2}$ reflected
across the difference vector $\mathbf{r}$. Due to the isotropy of
the system the two considered configurations have the same probability
to occur in bulk. Consequently, after performing the averages for
$\mathbf{J}_{\mathrm{vH}}^{\mathrm{\alpha,for}}$ in Eq.\ (\ref{eq:Jfor})
(or $\mathbf{J}_{\mathrm{vH}}^{\mathrm{\alpha,back}}$ in Eq.\ (\ref{eq:Jback})\hspace{0.03cm})
the contributions of $\mathbf{v}_{1}$ and $\mathbf{v}_{1}'$ perpendicular
to $\mathbf{r}$ (or the perpendicular contributions of $\mathbf{v}_{2}$
and $\mathbf{v}_{2}'$) cancel out and the van Hove
current is parallel to $\mathbf{r}$ (cf.\ Fig.~\ref{fig:sketch-cosy}).

Note that we use the same coordinate system for the backward and the
forward van Hove current with a radial unit vector pointing from $\mathbf{r}_{2}$
to $\mathbf{r}_{1}$. This convention implies that $J_{r}^{\mathrm{\alpha,back}}>0$
indicates a current of particle 2 at time $t_{2}$ towards
the position that would eventually be reached by particle 1 at $t_{1}$
as seen in Fig.\ \ref{fig:sketch-cosy}. However, $J_{r}^{\mathrm{\alpha,for}}>0$
would indicate a current of particle 1 at time $t_{1}$ away from
the position particle 2 had at $t_{2}$. Similar reasoning applies for  $J_{r}^{\mathrm{\alpha}}<0$.
For the diagonal components of $\JJ^{\alpha}$ it is implied
that positive values indicate aligned motion whereas negative values indicate opposing motion.

The trace of $\JJ^{\alpha}$ in
spherical coordinates reads
\begin{equation}
\Tr\JJ^{\alpha}(\mathbf{r},\tau)=J_{rr}^{\alpha}(r,\tau)+2J_{tt}^{\alpha}(r,\tau).\label{eq:Trace-spher}
\end{equation}
Hence, in a simulation all non-zero components of $\JJ^{\alpha}$
can be sampled without computing the transversal unit vectors $\mathbf{e}_{\varphi}$
and $\mathbf{e}_{\theta}$. The radial-radial component reads
\begin{align}
J_{rr}^{\mathrm{self}}(r,\tau)=\left\langle \sum_{i=1}^{N}\right. & \mathbf{e}_{r}\cdot\mathbf{v}_{i}(t_{1})\mathbf{e}_{r}\cdot\mathbf{v}_{i}(t_{2})\label{eq:Jrr-s}\\
& \times\delta(\mathbf{r}_{1}-\mathbf{r}_{i}(t_{1}))\delta(\mathbf{r}_{2}-\mathbf{r}_{i}(t_{2}))\Biggr\rangle ,\nonumber
\end{align}
\begin{align}
J_{rr}^{\mathrm{dist}}(r,\tau)=\left\langle \sum_{i=1}^{N}\sum_{j\neq i}^{N}\right.&\mathbf{e}_{r}\cdot\mathbf{v}_{i}(t_{1})\mathbf{e}_{r}\cdot\mathbf{v}_{j}(t_{2})\label{eq:Jrr-d}\\
&\times\delta(\mathbf{r}_{1}-\mathbf{r}_{i}(t_{1}))\delta(\mathbf{r}_{2}-\mathbf{r}_{j}(t_{2}))\Biggr\rangle,\nonumber
\end{align}
where, as before, $\mathbf{e}_{r}\equiv(\mathbf{r}_{1}-\mathbf{r}_{2})/\left|\mathbf{r}_{1}-\mathbf{r}_{2}\right|$.
The trans\-versal-transversal component can be obtained by combining
Eqs.\ (\ref{eq:Trace}), (\ref{eq:Trace-spher}), together with (\ref{eq:Jrr-s})
or (\ref{eq:Jrr-d}), which yields
\begin{align}
J_{tt}^{\mathrm{self}}(r,\tau)=&\,\frac{1}{2}\left\langle \sum_{i=1}^{N}\right.[\mathbf{v}_{i}(t_{1})\cdot\mathbf{v}_{i}(t_{2})-\mathbf{e}_{r}\cdot\mathbf{v}_{i}(t_{1})\mathbf{e}_{r}\cdot\mathbf{v}_{i}(t_{2})]\nonumber\\
&\times\delta(\mathbf{r}_{1}-\mathbf{r}_{i}(t_{1}))\delta(\mathbf{r}_{2}-\mathbf{r}_{i}(t_{2}))\Biggr\rangle,
\end{align}
\begin{align}
J_{tt}^{\mathrm{dist}}(r,\tau)=&\,\frac{1}{2}\left\langle \sum_{i=1}^{N}\sum_{j\neq i}^{N}\right.[\mathbf{v}_{i}(t_{1})\cdot\mathbf{v}_{j}(t_{2})\nonumber\\
&-\mathbf{e}_{r}\cdot\mathbf{v}_{i}(t_{1})\mathbf{e}_{r}\cdot\mathbf{v}_{j}(t_{2})]\\
&\times\delta(\mathbf{r}_{1}-\mathbf{r}_{i}(t_{1}))\delta(\mathbf{r}_{2}-\mathbf{r}_{j}(t_{2}))\Biggr\rangle.\nonumber
\end{align}

\subsection{Continuity relations\label{sub:CEq-theory}}

The hierarchy of the four two-body correlation functions (\ref{eq:rho2})
-- (\ref{eq:J2}) are related by five continuity equations. On the
scalar level there are relations between the temporal derivative of
the two-body density and the divergence of the forward and of the
backward van Hove current, 
\begin{equation}
\partial_{\tau}\rho_{2}^{\alpha}(r,\tau)=-\nabla\cdot\mathbf{J}_{\mathrm{vH}}^{\mathrm{\alpha,for}}(\mathbf{r},\tau),\label{eq:ceq1-1}
\end{equation}
\begin{equation}
\partial_{\tau}\rho_{2}^{\alpha}(r,\tau)=-\nabla\cdot\mathbf{J}_{\mathrm{vH}}^{\mathrm{\alpha,back}}(\mathbf{r},\tau).\label{eq:ceq2-1}
\end{equation}
These equations are derived as follows from the continuity equation for the density
operator and the current operator \citep{hansen_mcdonald},
\begin{equation}
\partial_{t_{1}}\hat{\rho}(1)=-\nabla_{1}\cdot\hat{\mathbf{J}}(1).\label{eq:op-ceq}
\end{equation}
After multiplying this relation
with $\hat{\rho}(2)$ and performing an average, one obtains
\begin{equation}
\bigl\langle\partial_{t_{1}}\hat{\rho}(1)\hat{\rho}(2)\bigr\rangle=-\bigl\langle\nabla_{1}\cdot\hat{\mathbf{J}}(1)\hat{\rho}(2)\bigr\rangle.\label{eq:ceq-derivation}
\end{equation}
The partial derivatives $\partial_{t_{1}}$ and $\nabla_{1}$ can
now be taken out of the average and be replaced by $\partial_{\tau}$
and $\nabla$, using Eqs.\ (\ref{eq:dt})
and (\ref{eq:dr}), which yields
\begin{equation}
\partial_{\tau}\rho_{2}(r,\tau)=-\nabla\cdot\mathbf{J}_{\mathrm{vH}}^{\mathrm{for}}(\mathbf{r},\tau).
\end{equation}
This equation can be split into its self and distinct part in analogy to Eq.\ (\ref{eq:separation-self-dist}), yielding Eq.\ (\ref{eq:ceq1-1}).
Equation (\ref{eq:ceq2-1}) can be derived analogously using exchanged space-time points
1 and 2.

On the vectorial level there are relations between the temporal derivatives
of the forward and of the backward van Hove current and the divergence
of the current-current correlator,
\begin{equation}
\,\partial_{\tau}\mathbf{J}_{\mathrm{vH}}^{\mathrm{\alpha,back}}(\mathbf{r},\tau)=-\nabla\cdot\JJ^{\alpha}(\mathbf{r},\tau).\label{eq:ceq4-1}
\end{equation}
\begin{equation}
\partial_{\tau}\mathbf{J}_{\mathrm{vH}}^{\mathrm{\alpha,for}}(\mathbf{r},\tau)=-\nabla\cdot\JJ^{\alpha}(\mathbf{r},\tau),\label{eq:ceq3-1}
\end{equation}
Equation (\ref{eq:ceq4-1}) is obtained by multiplying Eq.\ (\ref{eq:op-ceq}) with $\hat{\mathbf{J}}(2)$
and splitting into self and distinct parts. Equation (\ref{eq:ceq3-1}) is obtained analogously with exchanged space-time points 1 and 2.
Note that Eqs.\ (\ref{eq:ceq4-1}) and (\ref{eq:ceq3-1}) imply that the backward and the forward van
Hove current have the same numerical value in equilibrium, although
they describe different processes (cf.\ Sec.\ \ref{sub:Symm-and-Cosy}).
Therefore, we will drop the forward and backward label in those cases, and use the notation
\begin{equation}
\mathbf{J}_{\mathrm{vH}}^{\alpha}(\mathbf{r},\tau)\equiv\mathbf{J}_{\mathrm{vH}}^{\mathrm{\alpha,for}}(\mathbf{r},\tau)=\mathbf{J}_{\mathrm{vH}}^{\mathrm{\alpha,back}}(\mathbf{r},\tau).\label{eq:for-back}
\end{equation}
Consequently, Eqs.\ (\ref{eq:ceq1-1}) and (\ref{eq:ceq2-1}) and
Eqs.\ (\ref{eq:ceq3-1}) and (\ref{eq:ceq4-1}) are equivalent in
equilibrium. Equation (\ref{eq:for-back}) was also confirmed by sampling
$\mathbf{J}_{\mathrm{vH}}^{\mathrm{\alpha,for}}$ and $\mathbf{J}_{\mathrm{vH}}^{\mathrm{\alpha,back}}$
individually. The presented results for the radial component in Secs.\ \ref{sec:Results}
and \ref{sub:CEq-numerical} are the arithmetic mean of the backward
and forward current $J_{r}^{\alpha}=(J_{r}^{\mathrm{\alpha,back}}+J_{r}^{\mathrm{\alpha,for}})/2$
(cf.\ Eqs.\ (\ref{eq:JvH-radial}) and (\ref{eq:for-back})\hspace{0.03cm}),
again to reduce noise of the data.

For the second temporal derivative of the two-body density and the
tensor divergence of the current-current correlator we obtain
\begin{equation}
\partial_{\tau}\partial_{\tau}\rho_{2}^{\alpha}(r,\tau)=\nabla\nabla\mathbf{:}\JJ^{\alpha}(\mathbf{r},\tau)\label{eq:ceq5-1}
\end{equation}
by combining Eqs.\ (\ref{eq:ceq1-1})
and (\ref{eq:ceq3-1}) (or (\ref{eq:ceq2-1}) and (\ref{eq:ceq4-1})\hspace{0.03cm}).

The divergence on the right hand side of each continuity equation
can be calculated from the spherical components according to Eqs.\ (\ref{eq:Del*v}),
(\ref{eq:del*T}), and (\ref{eq:DelDel**T}), as deduced in the appendix
in Sec.\ \ref{sub:Del*JvH}. We verified the consistency of the
numerical results for the correlation functions using the continuity
equations (\ref{eq:ceq1-1}), (\ref{eq:ceq3-1}), and (\ref{eq:ceq5-1})
(see Sec.\ \ref{sub:CEq-numerical} for details).

\subsection{Static and dynamic test particle limit\label{sub:TPL}}

In the static case, $\tau=0$, the two-body density is related to the RDF
$g(r)$ \citep{hansen_mcdonald} via
\begin{equation}
\rho_{2}(r,0)=\rho\delta(\mathbf{r})+\rho^{2}g(r),\label{eq:RDF}
\end{equation}
where the delta distribution corresponds to the self part and the
RDF corresponds to the distinct part of $\rho_{2}$. In the static
TPL the RDF (multiplied by $\rho$) is identified with the one-body
density of a system exposed to an external potential \citep{STPL}.
The external potential has the form of the pair potential $u(r)$
where $r$ denotes the distance to the origin of a fixed coordinate
system. Hence, the external potential acts as a fixed particle --
denoted as test particle. As long as the test particle is kept fixed this system is in equilibrium and its  equilibrium density is
the same as the average of the instantaneous particle density around
an arbitrarily chosen particle in bulk.

The dynamic TPL pictures $\rho_{2}^{\alpha}$ as well as $\mathbf{J}_{\mathrm{vH}}^{\mathrm{\alpha,for}}$
as functions describing the relaxation of a non-equilibrium system
\citep{dTPL}. The initial state is the static test particle situation
with the test particle fixed at the origin. The test particle is
released at time $\tau=0$ and starts to diffuse away. The time dependent
one-body density and current (for the test particle and for the fluid
of all other particles) are then proportional to the two-body density
and current in the equilibrium bulk system. The contributions of
the test particle correspond to the self part (where particle $i=1$
is chosen as the test particle),
\begin{equation}
\rho_{\mathrm{tp}}^{\mathrm{self}}(r,\tau)\equiv\left\langle \delta(\mathbf{r}-\mathbf{r}_{1}(\tau))\right\rangle =\rho_{2}^{\mathrm{self}}(r,\tau)/\rho,\label{eq:rho-noneq-2}
\end{equation}
\begin{equation}
\mathbf{J}_{\mathrm{tp}}^{\mathrm{self}}(\mathbf{r},\tau)\equiv\left\langle \mathbf{v}_{1}(\tau)\delta(\mathbf{r}-\mathbf{r}_{1}(\tau))\right\rangle =\mathbf{J}_{\mathrm{vH}}^{\mathrm{self,for}}(\mathbf{r},\tau)/\rho.
\end{equation}
Recall that $\mathbf{r}_{1}(\tau)$, with time argument, denotes the
coordinates of particle 1 at time $\tau$. The functions of the other
$N-1$ particles correspond to the distinct parts,
\begin{equation}
\rho_{\mathrm{tp}}^{\mathrm{dist}}(r,\tau)\equiv\left\langle \sum_{i=2}^{N}\delta(\mathbf{r}-\mathbf{r}_{i}(\tau))\right\rangle =\rho_{2}^{\mathrm{dist}}(r,\tau)/\rho,\label{eq:rho-noneq-1-1}
\end{equation}
\begin{equation}
\mathbf{J}_{\mathrm{tp}}^{\mathrm{dist}}(\mathbf{r},\tau)\equiv\left\langle \sum_{i=2}^{N}\mathbf{v}_{i}(\tau)\delta(\mathbf{r}-\mathbf{r}_{i}(\tau))\right\rangle =\mathbf{J}_{\mathrm{vH}}^{\mathrm{dist,for}}(\mathbf{r},\tau)/\rho.\label{eq:JvH-DTPL}
\end{equation}
Note that we only consider the case of the forward van Hove current, such that a density (not current) operator acts at the earlier time.
For the inhomogeneous one-body functions in the test particle situation
a continuity equation can be obtained from Eq.\ (\ref{eq:ceq1-1})
or by performing an average of the
continuity equation for the one-body operators (\ref{eq:op-ceq}),
\begin{equation}
\partial_{\tau}\rho_{\mathrm{tp}}^{\alpha}(r,\tau)=-\nabla\cdot\mathbf{J}_{\mathrm{tp}}^{\alpha}(\mathbf{r},\tau).\label{eq:ceq-TPL}
\end{equation}

In order to obtain the BD results presented below, instead of explicitly implementing the test particle procedure, we rather
 sampled the dynamics of the system during the equilibrium time evolution. The test
 particle theory is, however, conceptually important for the
 splitting of the total force density into an adiabatic and a
 superadiabatic contribution.

\subsection{Construction of the adiabatic state\\and superadiabatic forces\label{sub:Superadiabatic-Force}}

Within the adiabatic approximation the two-body correlations are assumed to relax faster than the evolution of the dynamic density profiles.
Thus, the state of a system at any given time can be interpreted as an equilibrium state
and the interparticle forces (which are determined by the two-body correlations)
can be calculated from equilibrium statistical mechanics, as implemented in DDFT \citep{Archer-Evans04}.
The difference between the approximated adiabatic two-body correlations
and the full non-equilibrium two-body correlations yields an additional
contribution which is called the superadiabatic force \citep{Fortini-et-al,PFT13}.
We aim to calculate the adiabatic force and thereby the adiabatic
van Hove current by explicitly constructing the adiabatic state. Therefor
we first map the dynamic two-body density of the equilibrium bulk system to
a non-equilibrium one-body density via the dynamic TPL (see Sec.\ \ref{sub:TPL}).
Then we construct the adiabatic state by adjusting its equilibrium
density via an external potential as described in the following.

In BD the one-body current
is related to an internal force $\mathbf{F}_{\mathrm{int}}^{\alpha}(\mathbf{r},\tau)$
via the average velocity of the particles,
\begin{equation}
\mathbf{v}_{\mathrm{tp}}^{\alpha}(\mathbf{r},\tau)\equiv\frac{\mathbf{J}_{\mathrm{tp}}^{\alpha}(\mathbf{r},\tau)}{\rho_{\mathrm{tp}}^{\alpha}(r,\tau)}=\frac{\mathbf{J}_{\mathrm{vH}}^{\mathrm{\alpha,for}}(\mathbf{r},\tau)}{\rho_{2}^{\alpha}(r,\tau)}.\label{eq:vvH-1}
\end{equation}
Here the second equality follows from the relations between the one-body
functions in the TPL and the equilibrium two-body correlation functions,
(\ref{eq:rho-noneq-2}) -- (\ref{eq:JvH-DTPL}). The relaxation velocity
is related to the total internal force in analogy to Eq.\ (\ref{eq:overdamp}),
\begin{equation}
\mathbf{F}_{\mathrm{int}}^{\alpha}(\mathbf{r},\tau)=\xi\mathbf{v}_{\mathrm{tp}}^{\alpha}(\mathbf{r},\tau).\label{eq:F-v}
\end{equation}
Note that Eq.\ (\ref{eq:vvH-1}) implies that the internal force acting on the test particle system and the equilibrium system are equal.
Plugging in Eqs.~(\ref{eq:F-v}) and (\ref{eq:vvH-1})
into the continuity equation (\ref{eq:ceq-TPL}) yields
\begin{equation}
\partial_{\tau}\rho_{\mathrm{tp}}^{\alpha}(r,\tau)=-\xi^{-1}\nabla\cdot[\rho_{\mathrm{tp}}^{\alpha}(r,\tau)\mathbf{F}_{\mathrm{int}}^{\alpha}(\mathbf{r},\tau)].\label{time-evolution}
\end{equation}
The evolution of the density profile can be obtained from $\mathbf{F}_{\mathrm{int}}^{\alpha}(\mathbf{r},\tau)$ by integrating Eq.~(\ref{time-evolution}) in time.

In methods that use an adiabatic assumption, such as DDFT,
$\mathbf{F}_{\mathrm{int}}^{\alpha}(\mathbf{r},\tau)$ is approximated
by an adiabatic force $\mathbf{F}_{\mathrm{ad,\tau}}^{\mathrm{\alpha}}(\mathbf{r})$,
which can be defined as the force a system would experience in an
equilibrium state with the same instantaneous particle density as
the non-equilibrium system. The difference between $\mathbf{F}_{\mathrm{int}}^{\alpha}(\mathbf{r},\tau)$
and $\mathbf{F}_{\mathrm{ad,\tau}}^{\alpha}(\mathbf{r})$ is called
the superadiabatic force $\mathbf{F}_{\mathrm{sup}}^{\alpha}(\mathbf{r},\tau)$
and defined by splitting $\mathbf{F}_{\mathrm{int}}^{\alpha}(\mathbf{r},\tau)$
via
\begin{equation}
\mathbf{F}_{\mathrm{int}}^{\alpha}(\mathbf{r},\tau)\equiv\mathbf{F}_{\mathrm{ad,\tau}}^{\alpha}(\mathbf{r})+\mathbf{F}_{\mathrm{sup}}^{\alpha}(\mathbf{r},\tau).\label{eq:Fsup}
\end{equation}

We construct the adiabatic state following Ref.\ \citep{Fortini-et-al}
and calculate the internal adiabatic force via its external counter force
in the equilibrium situation. For each considered value of $\tau$
we simulate an equilibrium system with $N$ particles, referred to
as the adiabatic system. Note that in this system $\tau$ is not a real
time, but a label that specifies the reference state of the test particle
system. One particle (again we chose particle 1) is tagged and denoted
as the self particle. The other $N-1$ particles are denoted as distinct
particles. The equilibrium self and distinct one-body densities in
the adiabatic system, 
\begin{equation}
\rho_{\mathrm{ad,\tau}}^{\mathrm{self}}(r)\equiv\left\langle \delta(\mathbf{r}-\mathbf{r}_{1}(t))\right\rangle,
\end{equation}
\begin{equation}
\rho_{\mathrm{ad,\tau}}^{\mathrm{dist}}(r)\equiv\left\langle \sum_{i=2}^{N}\delta(\mathbf{r}-\mathbf{r}_{i}(t))\right\rangle,
\end{equation}
are obtained for convenience also in BD simulations (although a static method, such as Monte Carlo would suffice.)
The self particle and the distinct particles interact with self and distinct external adiabatic potentials
$V_{\mathrm{ad,\tau}}^{\mathrm{self}}(r)$ and $V_{\mathrm{ad,\tau}}^{\mathrm{dist}}(r)$, respectively
(which are not to be confused with the external potential that creates the static TPL in the previous section).
The external adiabatic potential is adjusted in an iterative procedure,
such that the density in the adiabatic system equals the one-body
density of the test particle system,
\begin{equation}
\rho_{\mathrm{ad,\tau}}^{\alpha}(r)=\rho_{\mathrm{tp}}^{\alpha}(r,\tau).\label{eq:rho-ad-tp-alpha}
\end{equation}
After an initial guess of $V_{\mathrm{ad,\tau}}^{\mathrm{\alpha,}(0)}(r)$,
we obtain $\rho_{\mathrm{ad,\tau}}^{\alpha,(0)}(r)$. In each iteration
step $n$ the external potential is adjusted locally according to
\begin{equation}
V_{\mathrm{ad,\tau}}^{\mathrm{\alpha,}(n)}(r)=V_{\mathrm{ad,\tau}}^{\mathrm{\alpha,}(n-1)}(r)+k_{\mathrm{B}}T\ln\frac{\rho_{\mathrm{ad,\tau}}^{\alpha,(n-1)}(r)}{\rho_{\mathrm{tp}}^{\alpha}(r,\tau)},\label{eq:Vad-iteration}
\end{equation}
where $\rho_{\mathrm{ad,\tau}}^{\alpha,(n-1)}(r)$ denotes the equilibrium
density obtained in the $(n-1)$th iteration step with the external
potential $V_{\mathrm{ad,\tau}}^{\mathrm{\alpha,}(n-1)}(r)$. In
practice we use BD simulations with $N=500$ particles and $2\times10^{6}$
samples in order to calculate the adiabatic density according to a
canonical average. The sampling of $\rho_{\mathrm{ad,\tau}}^{\alpha,(n-1)}(r)$
and the iteration of Eq.\ (\ref{eq:Vad-iteration})
are repeated 10 times.

In the static case, $\tau=0$, the adiabatic potential is known, as
this case is identical to the static TPL. The self adiabatic potential
$V_{\mathrm{ad,0}}^{\mathrm{self}}(r)$ is an infinitely deep and
infinitely narrow potential well that fixes the self particle at the
origin and thus generates the static TPL. The distinct adiabatic potential
$V_{\mathrm{ad,0}}^{\mathrm{dist}}(r)$ vanishes in this case. We
carried out the iteration as a consistency check for the method. Hereby
we represented the infinitely deep potential well of $V_{\mathrm{ad,0}}^{\mathrm{self}}(r)$
by keeping the coordinates of the self particle constant at $\mathbf{r}_{1}(t)=0$.
$V_{\mathrm{ad,0}}^{\mathrm{dist}}(r)$ was adjusted as in the dynamic
case and was found to vanish for all distances as expected.

The adiabatic system at each value of $\tau$ has a stationary equilibrium density distribution.
Therefore, the external adiabatic force (given by
the negative gradient of the external adiabatic potential $V_{\mathrm{ad,\tau}}^{\alpha}(r)$)
that generates the inhomogeneous density distribution balances the
internal adiabatic force $\mathbf{F}_{\mathrm{ad,\tau}}^{\alpha}(\mathbf{r})$,
that would drive the relaxation if the external field was switched
off, via
\begin{equation}
\mathbf{F}_{\mathrm{ad,\tau}}^{\alpha}(\mathbf{r})-\nabla V_{\mathrm{ad,\tau}}^{\alpha}(r)=0.\label{eq:balance}
\end{equation}
The adiabatic van Hove current then reads
\begin{equation}
\mathbf{J}_{\mathrm{ad}}^{\mathrm{\alpha}}(\mathbf{r},\tau)=\xi^{-1}\rho_{2}^{\alpha}(r,\tau)\nabla V_{\mathrm{ad,\tau}}^{\alpha}(r)\label{eq:Jad-theory}
\end{equation}
 (in analogy to Eqs.\ (\ref{eq:vvH-1})
and (\ref{eq:F-v})\hspace{0.03cm}). The superadiabatic van Hove current is obtained as the difference between
the adiabatic and the total van Hove current:
\begin{equation}
\mathbf{J}_{\mathrm{sup}}^{\mathrm{\alpha}}(\mathbf{r},\tau)=\mathbf{J}_{\mathrm{vH}}^{\mathrm{\alpha}}(\mathbf{r},\tau)-\mathbf{J}_{\mathrm{ad}}^{\mathrm{\alpha}}(\mathbf{r},\tau).\label{eq:Jsup-theory}
\end{equation}

\subsection{Freely relaxing reference system\label{sub:Relaxing-System}}

In order to assess the dynamical properties of the LJ liquid we examine a reference system,
that is equilibrated like the LJ system described in Sec.\ \ref{sub:Brownian-Dynamics},
but the particle-particle interactions are switched off at time $t=0$.
Then we calculate the two-body correlation functions for $t_{2}=0$ as functions of $\tau=t_{1}$.
Due to the switch off the correlations decay via free diffusion (as is characteristic of the ideal gas).
For $\tau=0$ and for $\tau\rightarrow\infty$ the two-body density of this system equals the one of the equilibrium LJ system,
which makes the system a useful reference for assessing the influence of the internal interactions on the dynamic decay.
The spatial symmetries described in Sec.\ \ref{sub:Symm-and-Cosy} still apply.
However, the switch off creates a non-equilibrium situation and the temporal symmetries do no longer apply for the distinct parts.
In particular the symmetry between the forward and backward van Hove currents is lost.

The self parts of the correlation functions in this system equal the ones of an ideal gas.
The choice of the random force in Eq.\ (\ref{eq:ran_Force-1}) implies that the
random displacement vector (without particle-particle interactions)
$\Delta\mathbf{r}_{i}=\dot{\mathbf{r}}_{i}(t)\Delta t$ is also Gaussian
distributed. Furthermore the variance
of the displacement vector after a time $\tau$, $\sigma_{\tau}^{2}$,
(i.e.{\ }after $\tau/\Delta t$ time steps)
reads
\begin{equation}
\sigma_{\tau}^{2}=\frac{\tau}{\Delta t}\sigma_{\Delta t}^{2}=\tau\Delta t\sigma_{v}^{2}=\frac{\tau\Delta t\sigma_{R}^{2}}{\xi^{2}}=\frac{2\tau k_{\mathrm{B}}T}{\xi}=2D_{0}\tau.\label{eq:Var}
\end{equation}
Here the variance of the displacement vector after one time step,
$\sigma_{\Delta t}^{2}$, and the variance of the displacement velocity,
$\sigma_{v}^{2}$, are linked to $\sigma_{R}^{2}$ via Eqs.{\ }(\ref{eq:trajectory-1})
and (\ref{eq:overdamp}) and $D_{0}=k_{\mathrm{B}}T/\xi$ is the
free diffusion coefficient \citep{ES}. The self two-body density is proportional
to the probability distribution of displacement vectors,
\begin{equation}
\rho_{2}^{\mathrm{self}}(r,\tau)=\begin{cases}
\rho\delta(\mathbf{r}) & \tau=0\\
\rho(2\pi\sigma_{\tau}^{2})^{-3/2}\exp\left(-\frac{r^{2}}{2\sigma_{\tau}^{2}}\right) & \tau>0.
\end{cases}\label{eq:rho2-self-re}
\end{equation}
The delta distribution at $\tau=0$ is given by Eq.\ (\ref{eq:RDF}) or can be calculated by taking the limit $\tau\rightarrow0$
of the Gaussian distribution in the second line. The prefactors are determined by the normalization condition (\ref{eq:norm-self}).
Also note that Eq.\ (\ref{eq:rho2-self-re}) is the solution of the free diffusion equation $\partial_\tau\rho_{2}(r,\tau)=D_0 \nabla^2\rho_{2}(r,\tau)$
with the initial condition $\rho_{2}(r,0)=\rho\delta(\mathbf{r})$.

For ideal motion the average in Eq.\ (\ref{eq:Jfor}) or (\ref{eq:J2})
simplifies to integrals over all random displacement velocities weighted
with their probability distributions. Hence, $\mathbf{J}_{\mathrm{vH}}^{\mathrm{self}}$
and $\JJ^{\mathrm{self}}$ can
be calculated analytically. This calculation is carried out in detail
in the appendix in Sec.\ \ref{sub:J-self-re} for $\tau>0$ and yields
\begin{equation}
\mathbf{J}_{\mathrm{vH}}^{\mathrm{self}}(\mathbf{r},\tau)=\begin{cases}
0 & \tau=0\\
\frac{\mathbf{r}}{2\tau}\rho_{2}^{\mathrm{self}}(r,\tau) & \tau>0,
\end{cases}\label{eq:JvH-self-re}
\end{equation}
\begin{equation}
\JJ^{\mathrm{self}}(\mathbf{r},\tau)=\begin{cases}
\frac{2\rho k_{\mathrm{B}}T}{\xi}\delta(\tau)\delta(\mathbf{r})\underline{\underline{\mathbf{I}}} & \tau=0\\
\frac{1}{4\tau^{2}}\left(\mathbf{r}\mathbf{r}-\sigma_{\tau}^{2}\underline{\underline{\mathbf{I}}}\right)\rho_{2}^{\mathrm{self}}(r,\tau) & \tau>0.
\end{cases}\label{eq:J2-self-re}
\end{equation}
The van Hove current at time $\tau=0$ vanishes due to the isotropy in
the system. The current-current correlator at time zero is determined
by the self correlation of the Brownian forces (\ref{eq:corr-force}).
In the spherical coordinate system $\JJ^{\mathrm{self}}$
is a diagonal tensor. The radial-radial component and the two equal
transversal-transversal components at non-zero times are given, respectively,
by
\begin{equation}
J_{rr}^{\mathrm{self}}(r,\tau)=\frac{r^{2}-\sigma_{\tau}^{2}}{4\tau^{2}}\rho_{2}^{\mathrm{self}}(r,\tau),\label{eq:J-rr-self-re}
\end{equation}
\begin{equation}
J_{tt}^{\mathrm{self}}(r,\tau)\equiv J_{\theta\theta}^{\mathrm{self}}(r,\tau)=J_{\varphi\varphi}^{\mathrm{self}}(r,\tau)=-\frac{\sigma_{\tau}^{2}}{4\tau^{2}}\rho_{2}^{\mathrm{self}}(r,\tau).\label{eq:J-tt-self-re}
\end{equation}

It is laborious but straightforward to show that the analytical solutions for the ideal two-body correlation functions
satisfy the continuity equations (\ref{eq:ceq1-1}), (\ref{eq:ceq3-1}), and (\ref{eq:ceq5-1}).
Note that the van Hove current has only one non-zero (radial) component and could also be calculated
by integrating the continuity equation (\ref{eq:ceq1-1}). The current-current
correlator however has two independent components. Therefore, it can
not be obtained from the continuity equation in a simple way.

\subsection{Asymptotic expansion of the two-body density\label{sub:log-rho}}

In fluids the RDF approaches a bulk value of unity for $r\rightarrow\infty$.
Fisher and Widom investigated the asymptotic decay of $g(r)$
in one dimension \citep{FW}. Further research was carried out for
three dimensional models and different kinds of inhomogeneous situations
(see e.g.\ \citep{Henderson-Sabeur92,FW-HSAY}). The following presentation
follows closely Ref.\ \citep{Dijkstra-Evans00}. For short-ranged
pair potentials, such as the truncated and shifted LJ potential used
in this work, the deviation of the RDF from the bulk value can be
expanded in a sum of exponentials, 
\begin{equation}
\frac{r}{\sigma}(g(r)-1)=\sum_{n}A_{n}e^{iq_{n}r}.\label{eq:Expansion}
\end{equation}

Here $r/\sigma$ is a geometrical factor for three-dimensional
systems, $q_{n}$ are the solutions with positive imaginary part of
\begin{equation}
1-\rho\tilde{c}(q_{n})=0\label{poles}
\end{equation}
with Fourier transform of the direct correlation function, $\tilde{c}(q)$,
and
\begin{equation}
A_{n}=-\frac{q_{n}}{2\pi\sigma\rho^{2}}\left(\frac{\mathrm{d}\tilde{c}}{\mathrm{d}q}(q_{n})\right)^{-1}
\end{equation}
are the amplitudes of the components. In general there is an infinite number of solutions for Eq.~(\ref{poles})
and therefore Eq.\ (\ref{eq:Expansion}) is a series.
The ultimate long-range decay of $g(r)$ is then determined by the
component(s) that possess the slowest exponential decay, i.e.\ the
term(s) with the smallest imaginary part of $q_{n}$ in Eq.\ (\ref{eq:Expansion}).
There are two possibilities; either this is purely imaginary, $q_{0}\equiv i\alpha_{0}$,
yielding monotonic exponential long-range decay,
\begin{equation}
\frac{r}{\sigma}(g(r)-1)\sim Ae^{-\alpha_{0}r},\label{eq:damped-mono}
\end{equation}
or those are a pair of conjugated complex numbers $\tilde{q}_{0}\equiv\pm\tilde{\alpha}_{1}+i\tilde{\alpha}_{0}$
yielding oscillatory exponentially damped decay,
\begin{equation}
\frac{r}{\sigma}(g(r)-1)\sim\tilde{A}e^{-\tilde{\alpha}_{0}r}\cos(\tilde{\alpha}_{1}r-\theta).\label{eq:damped-osci}
\end{equation}
The amplitudes $A$, $\tilde{A}$ and the phase $\theta$ can be
calculated explicitly from $\tilde{c}(q)$ (see e.g.\ \citep{Asymptotic1,Asymptotic2})
and are specific for the kind of inhomogeneity around $r=0$.
Here the inhomogeneity consists of the pair potential of the self particle (cf.\ Sec.\ \ref{sub:TPL}).
However, the inverse decay lengths $\alpha_{0}$, $\tilde{\alpha}_{0}$,
and the wave number of the periodicity, $\tilde{\alpha}_{1}$, are bulk properties
that only depend on the model and the state point of the fluid. One
expects \citep{FW} exponential decay for low densities and temperatures
and oscillatory damped decay for high densities and temperatures.
The corresponding regions in the phase diagram are separated by the
so called Fisher-Widom line. For the state point $T=0.80\epsilon/k_{\mathrm{B}}$
and $\rho=0.84/\sigma^3$ examined below, we expect oscillatory
damped decay. Hence, we base our further considerations on Eq.\ (\ref{eq:damped-osci}).

Combining Eqs.\ (\ref{eq:RDF}) and (\ref{eq:damped-osci})
and taking the logarithm of the absolute value, one obtains
\begin{align}
\ln\left(\frac{r}{\sigma}\left|\rho_{2}^{\mathrm{dist}}(r,0)/\rho^{2}-1\right|\right)\sim & -\tilde{\alpha}_{0}r+\ln\tilde{A}\label{eq:ln(r(rho-1))}\\
&+\ln\left|\cos(\tilde{\alpha}_{1}r-\theta)\right|.\nonumber
\end{align}
The absolute value is used because the cosine attains also negative values.
We calculate the left hand side of Eq.\ (\ref{eq:ln(r(rho-1))})
in the equilibrium bulk described in Sec.\ \ref{sub:Brownian-Dynamics}
with $N=6400$ particles. From the numerical value we obtain $\tilde{\alpha}_{0}$
and $\tilde{A}$, as $\ln\left|\cos(\tilde{\alpha}_{1}r-\theta)\right|$
is always negative and the straight line $-\tilde{\alpha}_{0}r+\ln\tilde{A}$
is the envelope of the right-hand side. The values
of $\tilde{\alpha}_{1}$ and $\theta$ can be obtained from the poles
of the function as these correspond to the zeros of the cosine.

We aim to extend this description of the long-range decay for dynamic
correlation functions by calculating the left-hand side of Eq.\ (\ref{eq:ln(r(rho-1))})
for non-zero times, i.e.\ $\ln\left(r/\sigma\left|\rho_{2}^{\mathrm{dist}}(r,\tau)/\rho^{2}-1\right|\right)$,
and by observing the decay of the correlations. In order to assess the underlying
mechanisms we also calculate the dynamic long-range decay in the freely
relaxing reference system (see Sec.\ \ref{sub:Relaxing-System}).
By comparing both systems we intend to determine the effect of the
particle-particle interactions on the decay of the correlations.

\section{Numerical results\label{sec:Results}}

\subsection{Self correlation functions\label{sub:Results-Self}}

\begin{figure*}[t]
\includegraphics[width=15cm,angle=0]{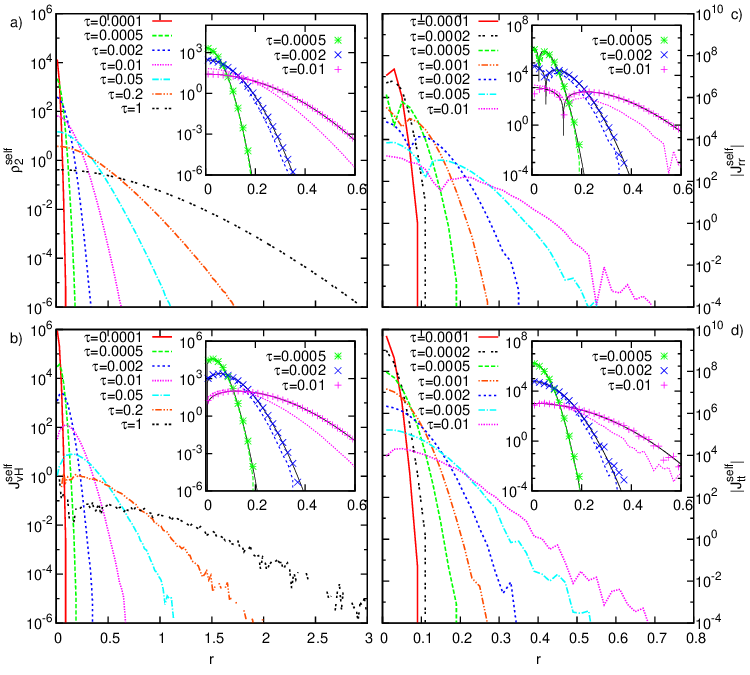}
\caption{\label{fig:ALL_self}Simulation results for the self two-body correlation functions plotted
on a logarithmic scale as functions of distance $r$ at time differences
$\tau$ as indicated. a) two-body density $\rho_{2}^{\mathrm{self}}$
and b) radial component of the van Hove current, $J_{r}^{\mathrm{self}}$,
at time differences from $\tau=0.0001$ to 1 as indicated. c)
absolute values of the radial-radial component, $\left|J_{rr}^{\mathrm{self}}\right|$,
and d) of the transversal-transversal component of the current-current
correlator, $\left|J_{tt}^{\mathrm{self}}\right|$, at time differences
from $\tau=0.0001$ to 0.01 as indicated. The insets show comparisons
between the simulation results in the LJ system (lines, as in main panels) and the simulation results of the freely relaxing reference
system (symbols) at $\tau=0.0005$, 0.002, and 0.01. Symbols of
the simulation results for the freely relaxing reference system lie
on top of the analytical solutions (black solid lines) according
to Eqs.\ (\ref{eq:rho2-self-re}), (\ref{eq:JvH-self-re}), and (\ref{eq:J2-self-re})
at each time.}
\end{figure*}

The non-zero components of the self parts of the two-body correlation functions, $\rho_{2}^{\mathrm{self}}$, $J_{r}^{\mathrm{self}}$,
$J_{rr}^{\mathrm{self}}$, and $J_{tt}^{\mathrm{self}}$ for $\tau>0$
are shown in Fig.\ \ref{fig:ALL_self} on logarithmic scales. (The case $\tau=0$ is discussed below.) For
the self two-body density $\rho_{2}^{\mathrm{self}}$ we obtained
a Gaussian-like distribution where the variance of the distribution
grows with increasing time difference $\tau$, as is known for diffusive
motion (cf.\ Eq.\ (\ref{eq:rho2-self-re})\hspace{0.03cm}). Our
simulation results are shown in Fig.\ \ref{fig:ALL_self}a). 

The radial component of the self  van Hove current, $J_{r}^{\mathrm{self}}$,
is shown in Fig.\ \ref{fig:ALL_self}b).
For small distances $J_{r}^{\mathrm{self}}$ approaches zero.
This indicates that particles close to their original position do not prefer a certain direction in their motion and reflects the isotropy of the system.
For greater distances $J_{r}^{\mathrm{self}}$ is permanently positive, which indicates motion towards higher distances
and is consistent with the broadening Gaussian distribution of the self two-body density (cf.\ Fig.\ \ref{fig:ALL_self}a)\hspace{0.03cm}).
The decay of $J_{r}^{\mathrm{self}}$ at large distances is caused by the small number of particles reaching
higher distances as seen in the decay of the self two-body density.

Figure \ref{fig:ALL_self}c) shows the absolute value of the radial-radial
component of the self current-current correlator, $J_{rr}^{\mathrm{self}}$.
For $\tau\neq0$ and small values of $r$ the values of $J_{rr}^{\mathrm{self}}$
are negative, which indicates opposing motion of the self particle
at the two considered times 0 and $\tau$. This indicates that particles
that are found near their original position after some time $\tau$
have moved some distance and turned around, returning to their original
position with oppositely directed velocity. For greater values of
$r$ the values of $J_{rr}^{\mathrm{self}}$ are positive. The positive
values arise from particles moving in the same direction at times
0 and $\tau$. The transversal-transversal component $J_{tt}^{\mathrm{self}}$
is a Gaussian-like distribution with negative prefactor. See Fig.\ \ref{fig:ALL_self}d)
for a plot of $J_{tt}^{\mathrm{self}}$ as a function of $r$.

The simulation results of the LJ system are compared to the ideal
dynamics of the freely relaxing reference system, as shown in the
insets of Fig.\ \ref{fig:ALL_self}. The simulation results of the
freely relaxing system are in good agreement with the results from
the analytic calculations (also shown in the insets) given by Eqs.\ (\ref{eq:rho2-self-re}),
(\ref{eq:JvH-self-re}), and (\ref{eq:J2-self-re}). This serves as
a consistency check e.g.\ for the correct magnitude of the random displacements
in the simulations. The qualitative behavior of each correlation
function in both the LJ and in the ideal systems is identical. For
$\tau<0.0002$ also the numerical values for both systems are
equal. For larger values of $\tau$ the correlation functions in the
LJ system broaden more slowly than in the freely relaxing system,
as can be seen in each inset of Fig.\ \ref{fig:ALL_self}a) -- d).
The dynamics shift from ideal-like short-time diffusion to long-time
diffusion with a reduced diffusion coefficient as the movement is
inhibited by the surrounding particles. For $\tau=0$ the simulation
results for the two-body correlation functions are not shown in the
figures, as the values are zero for $r\neq0$ and
show different singularities at $r=0$.
As the self correlation functions approach the ideal motion in the limit of small times
we assume that the singularities in the LJ system at $\tau=0$ are equal to the singularities of the ideal gas,
which are given by Eqs.\ (\ref{eq:rho2-self-re}), (\ref{eq:JvH-self-re}), and (\ref{eq:J2-self-re}).

\subsection{Distinct correlation functions\label{sub:Results-Distinct}}

\begin{figure}
\includegraphics[width=8.8cm,angle=0]{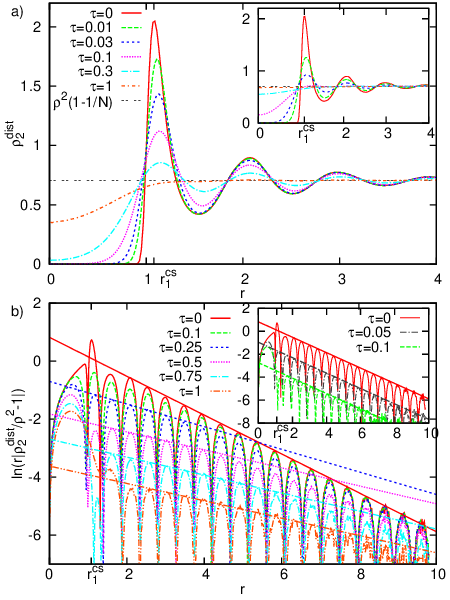}
\caption{\label{fig:rho2_dist} a) Distinct two-body density $\rho_{2}^{\mathrm{dist}}$ as function
of distance $r$ at time differences from $\tau=0$ to $1$ as indicated. The double dashed
lines at $\rho^{2}(1-1/N)$ correspond to the value of $\rho_{2}^{\mathrm{dist}}$
for $\tau\rightarrow\infty$. b) Absolute logarithmic deviation of the two-body density from the bulk value
(left hand side of Eq.\ (\ref{eq:ln(r(rho-1))})\hspace*{0.03cm})
as function of distance $r$ at time differences from $\tau=0$ to 1 as indicated.
The logarithmic deviation was obtained with $N=6400$ particles. The straight lines
serve as a guide to the eye for the maxima of the oscillations at
each time. Both insets show the same functions as the big plots obtained in the freely
relaxing reference system. $r_{1}^{\mathrm{cs}}$ indicates the
position of the first maximum of $\rho_{2}^{\mathrm{dist}}$.}
\end{figure}

The numerical results for the distinct two-body density $\rho_{2}^{\mathrm{dist}}$ are shown in Fig.~\ref{fig:rho2_dist}.
In the static case, i.e.\ for $\tau=0$, $\rho_{2}^{\mathrm{dist}}$
(see red solid line in Fig.\ \ref{fig:rho2_dist}a)\hspace{0.03cm}) or equivalently
the RDF (cf.\ Eq.\ (\ref{eq:RDF})\hspace{0.03cm}) shows the well-known
behavior characteristic of a dense liquid. The values of $\rho_{2}^{\mathrm{dist}}$
practically vanish for $r<0.85$ due to the repulsive cores of the particles.
For larger distances $\rho_{2}^{\mathrm{dist}}$
shows oscillatory behavior, approaching the bulk value of the two-body
density of $\rho_{2}^{\mathrm{dist}}(r\rightarrow\infty)=\rho^{2}$. The locations
of the peaks correspond to correlation shells around the self particle
with high local particle density. The corresponding positions of the
1st, 2nd, and 3rd maxima are $r_{1}^{\mathrm{cs}}=1.08$, $r_{2}^{\mathrm{cs}}=2.07$,
and $r_{3}^{\mathrm{cs}}=2.97$. The density oscillations result
from dominance of repulsive over attractive forces in the system \citep{FW}.
In other words, particles in the first correlation shell tend to arrange
closely around the self particle due to confinement by the surrounding
particles. Analogously particles in outer correlation shells arrange
closely around the inner correlation shells.
For increasing $\tau$ the correlation shells
decay and the depletion zone near $r=0$ is filled. The peaks decay much
faster in the freely relaxing system than in the LJ system (cf.\ inset
of Fig.\ \ref{fig:rho2_dist}a)\hspace{0.03cm}). As we will scrutinize below
not only the decay rate is different in each system but also the
mechanism. At large $\tau$ the two-body density approaches
a uniform distribution with $\rho_{2}^{\mathrm{dist}}(\tau\rightarrow\infty)=\rho^{2}(1-1/N)$
(also shown in the diagram). The factor $(1-1/N)$ is a finite size
effect arising from the normalization condition
for $\rho_{2}^{\mathrm{dist}}$, (\ref{eq:norm-dist}).

In order to reveal more detail we plotted the absolute deviation of the two-body density
from its bulk value on a logarithmic scale (see Fig.\ \ref{fig:rho2_dist}b)\hspace{0.03cm}),
as is usually used to examine the long-range behavior (cf.\ Sec.\ \ref{sub:log-rho}).
Note that maxima in this plot are equivalent to either maxima or minima of $\rho_{2}^{\mathrm{dist}}$
on the linear plot in Fig.\ \ref{fig:rho2_dist}a).
At $\tau=0$ (red solid curve) we obtain an exponentially decaying oscillating function for $r>2$.
The exponential decay is indicated by a linear envelope, which is also shown in the diagram.
The good agreement of the envelope with the maxima of the oscillations resembles the finding
that Eq.~(\ref{eq:ln(r(rho-1))}) is a very good approximation even for intermediate distances (see e.g.~\citep{Asymptotic2,Leote-de-Carvalho-Evans94}).
From the straight line we obtain a static inverse correlation length of $\tilde{\alpha}_{0}=0.67$
and an amplitude of $\ln\tilde{A}=0.82$. The envelope of the static two-body correlations then reads
\begin{equation}
-\tilde{\alpha}_{0}r+\ln \tilde{A}=-0.67r+0.82.\label{eq:static-envelope}
\end{equation}
From the zeros of the oscillations we obtain $\tilde{\alpha}_{1}=6.69$ and $\theta=1.3$.
Leote de Carvalho $et\,al.$ report values of  $\tilde{\alpha}_{0}=1.0041$ and $\tilde{\alpha}_{1}=6.4265$ for a LJ fluid truncated at $r_c=2.5$ at a state point with higher temperature and lower density of $T=1.2\epsilon/k_{\mathrm{B}}$
and $\rho=0.715/\sigma^3$ \citep{Leote-de-Carvalho-Evans94}.
The higher value of $\tilde{\alpha}_{0}$ represents a faster exponential decay of the correlations, which is caused by the lower density.
The lower value of $\tilde{\alpha}_{1}$ mirrors slower oscillations of the two-body density and is a consequence of the higher temperature.

For $\tau\neq0$ the oscillations of the two-body density persist and the correlations decay in
time, which is indicated by the peak heights decreasing. Hereby the
peaks start to decrease faster for small distances, whereas the correlations
at larger distances still persist. Thus, there are two distinct regions
with different behavior; an inner region with a new $dynamic$ inverse
correlation length $\tilde{\beta}_{0}(\tau)$ and decreasing amplitude
$\ln B(\tau)$ and an outer region where the correlations are unchanged
and remain equal to the static value.
We aligned straight lines to the peaks in the inner region (cf.\ Fig.\ \ref{fig:rho2_dist}b)\hspace{0.03cm})
to estimate values of $\tilde{\beta}_{0}(\tau)$ and $\ln B(\tau)$.
Between $\tau=0$ and 0.7 the value of $\tilde{\beta}_{0}(\tau)$
drops from the initial value $\tilde{\beta}_{0}(0)=\tilde{\alpha}_{0}=0.67$
to $\tilde{\beta}_{0}(0.7)\approx0.28$, which implies that the
dynamic correlation length is larger than the static correlation length.
For $\tau>0.7$ the inverse dynamic correlation length is approximately
constant at $\tilde{\beta}_{0}\approx0.28$ and the dynamic
amplitude $\ln B(\tau)$ decreases linearly in time and
follows approximately the law $\ln B(\tau)\equiv-B_{1}\tau+B_{2}\approx-4\tau+0.2$.
The envelope in the dynamic region then reads
\begin{equation}
-\tilde{\beta}_{0}r-B_{1}\tau+B_{2}\approx-0.28r-4\tau+0.2.\label{eq:dynamic-envelope}
\end{equation}
The boundary between the dynamic intermediate region and the static
long-range region could be measured with the distance at the intersection
of the two envelopes Eqs.\ (\ref{eq:static-envelope}) and (\ref{eq:dynamic-envelope}),
$r_{\mathrm{int}}$, which reads
\begin{equation}
r_{\mathrm{int}}=\frac{B_{1}}{\tilde{\alpha}_{0}-\tilde{\beta}_{0}}\tau+\frac{\ln A-B_{2}}{\tilde{\alpha}_{0}-\tilde{\beta}_{0}}\approx10\tau+1.6.\label{eq:Interaction-Speed}
\end{equation}
The prefactor of $\tau$, $v_{\mathrm{int}}\equiv B_{1}/(\tilde{\alpha}_{0}-\tilde{\beta}_{0})\approx10$,
has the dimension of a velocity and measures the speed of the dynamic
region expanding in the fluid.
Note that the crossover from dynamic to static region in our numerical data is not as sharp as Eq.\ (\ref{eq:Interaction-Speed}) suggests
and the obtained values of $r_{\mathrm{int}}$ and $v_{\mathrm{int}}$ are estimates that depend on the utilized procedure.

A possible explanation for this remarkable behavior can be based on the TPL.
Recall that in the static TPL the two-body density at $\tau=0$ is interpreted as a static equilibrium quantity, which means that
the oscillations of the static two-body density $\rho_{2}^{\mathrm{dist}}(r,0)$ are
in principle stable as long as the self particle stays fixed at its
position at the origin of the coordinate system. As pointed out before, the $n$th correlation shell is created by confinement
of its particles between the $(n-1)$th and the $(n+1)$th correlation
shell (or between the second correlation shell and the self particle
in case of the first correlation shell). When the self particle is
released at $\tau=0$ -- within the dynamic TPL description -- these correlation
shells are still in place and still stabilize each other. When the
self particle starts to diffuse away, the confinement of the first correlation
shell is broken up. Thus, the first correlation shell dissolves and
therefore no longer confines the second correlation shell. This effect
propagates throughout the system yielding the behavior observed in
the simulations. In this picture $v_{\mathrm{int}}\approx10$
is interpretable as the speed of the ``starting signal'' for the
decay propagating through the fluid.
The explanation is supported by the comparison with the freely relaxing reference system.
Here the switch off of the pair forces at $\tau=0$ instantaneously pushes the whole fluid out of equilibrium.
Consequently, $\rho_{2}^{\mathrm{dist}}$ does not exhibit the delayed decay of the peaks
but the correlations decay concurrently throughout the fluid (see inset of Fig.~\ref{fig:rho2_dist}b)\hspace{0.03cm}).

\begin{figure}
\includegraphics[width=8.8cm,angle=0]{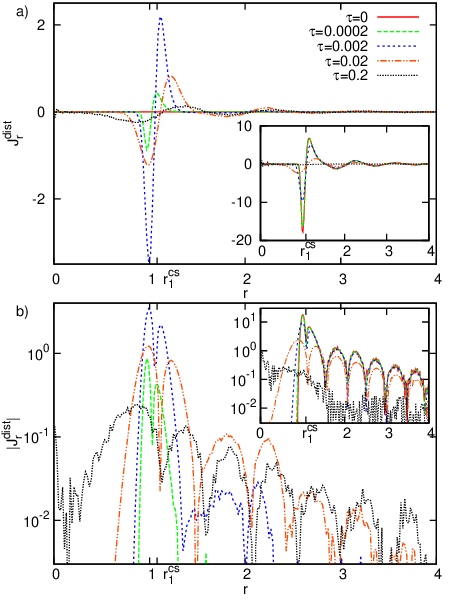}
\caption{\label{fig:JvH-dist-2}Radial component of the distinct van Hove current, $J_{r}^{\mathrm{dist}}$,
as function of distance $r$ at time differences from $\tau=0$
to 0.2 as indicated a) on linear a scale and b) absolute values on
a logarithmic scale. The insets show the radial component of the
forward van Hove current, $J_{r}^{\mathrm{dist,for}}$, in the freely
relaxing reference system. $r_{1}^{\mathrm{cs}}$
indicates the position of the first maximum of $\rho_{2}^{\mathrm{dist}}$.}
\end{figure}

The radial component of the distinct van Hove current, $J_{r}^{\mathrm{dist}}$,
is shown in Fig.\ \ref{fig:JvH-dist-2}a)
(linear scale) and \ref{fig:JvH-dist-2}b) (logarithmic scale). At $\tau=0$ the distinct van Hove current
vanishes for all distances, which is a direct consequence of the equilibrium
situation: In equilibrium there is detailed balance. Hence, the fraction
of distinct particles moving towards the self particle is balanced
by an equal fraction of particles moving away from the self particle.
For non-zero times a negative peak at $r\lesssim r_{1}^{\mathrm{cs}}$ and a positive
peak at $r\gtrsim r_{1}^{\mathrm{cs}}$
grow and broaden, and subsequently decay (see Fig.\ \ref{fig:JvH-dist-2}a)\hspace{0.03cm}).
A similar behavior is observable around $r\approx r_{2}^{\mathrm{cs}}$
and $r\approx r_{3}^{\mathrm{cs}}$. Yet these peaks reach their maximum
value at later times and have smaller maximum values than the nearest
neighbor peaks around $r\approx r_{1}^{\mathrm{cs}}$ (see Fig.\ \ref{fig:JvH-dist-2}b)\hspace{0.03cm}).

When viewing the results as representing the forward van Hove current
$\mathbf{J}_{\mathrm{vH}}^{\mathrm{dist,for}}$, the negative peaks
indicate particles leaving their correlation shell with negative radial
velocity towards smaller distance. The positive peaks arise from
particles that leave their correlation shells with a positive velocity
towards larger distance. This behavior is consistent with the filling
of the low density regions of the distinct two-body density (cf.\ Fig.\ \ref{fig:rho2_dist}a)\hspace{0.03cm}).
For the backward van Hove current $\mathbf{J}_{\mathrm{vH}}^{\mathrm{dist,back}}$
an analogous interpretation with reversed time applies.

The situation is different in the freely relaxing system. Here the backward van Hove current is zero for all times,
as the velocity of one particle at $\tau=0$ does not correlate with the positions of the other particles at later times.
The forward van Hove current (shown in the insets of Fig.\ \ref{fig:JvH-dist-2}a) and b)\hspace{0.03cm}) jumps instantaneously
from the static value $J_{r}^{\mathrm{dist,for}}=0$ to finite non-zero values and the peak values are larger than in the interacting system.
The instantaneous jump is caused by the switch off of the particle-particle interactions, which creates a non-equilibrium situation.
The behavior is consistent with the peaks of $\rho_{2}^{\mathrm{dist}}$ decreasing faster in the reference system than in the LJ system
and with the absence of the outer static region (cf.\ inset of Fig.\ \ref{fig:rho2_dist}b)\hspace{0.03cm}).
However, the signs of the peaks at corresponding distances are equal and the peaks are shaped similarly in both systems.
Hence, the same interpretation as for $\mathbf{J}_{\mathrm{vH}}^{\mathrm{dist,for}}$
given above for the LJ system applies for the freely relaxing system,
where the peaks correspond to particles leaving their correlation
shells towards a uniform density distribution.

\begin{figure}
\includegraphics[width=8.8cm,angle=0]{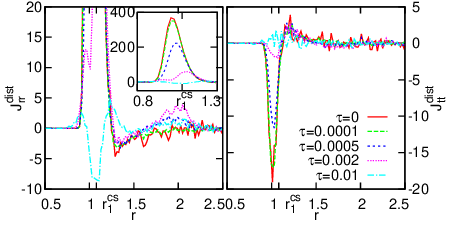}
\caption{\label{fig:J2-dist}Distinct current-current correlator, a) radial-radial component $J_{rr}^{\mathrm{dist}}$
and b) transversal-transversal component $J_{tt}^{\mathrm{dist}}$
as functions of distance $r$ at time differences from $\tau=0$
to 0.01 as indicated. The inset in diagram a) shows the nearest neighbor
peak of $J_{rr}^{\mathrm{dist}}$ around $r=1.02$, which is off scale in the main panel.
$r_{1}^{\mathrm{cs}}$ indicates the position of the first maximum
of $\rho_{2}^{\mathrm{dist}}$.}
\end{figure}

We next turn to the current-current correlator $\JJ^{\mathrm{dist}}$ which is shown in Fig.~\ref{fig:J2-dist}. 
The radial-radial component of the static distinct current-current
correlator, $J_{rr}^{\mathrm{dist}}$, (see Fig.~\ref{fig:J2-dist}a)\hspace{0.03cm})
has only a single prominent peak with a maximum value of $J_{rr}^{\mathrm{dist}}=374$
at $r=1.02$. The position of the maximum is smaller than $r_{1}^{\mathrm{cs}}$,
which indicates that the peak arises from two particles undergoing
a collision. An explanation is that at the turning point of the collision
(i.e.\ the time, when the distance is minimal) the particles have
an equal velocity in radial direction, which results in a strong positive correlation.
This notion is supported by the absence of a static peak at $r\approx2$.
The large nearest neighbor peaks of the static correlations decay within a time of $\tau\approx0.005$.
During this decay a negative peak grows within the decaying positive static peak.
The maximum amplitude of this dynamic nearest neighbor peak of $J_{rr}^{\mathrm{dist}}\approx-11$
is attained at $\tau\approx0.005$. Also a positive peak at $r\approx2$,
arising from particles in the second correlation shell, grows and
attains its maximum of $J_{rr}^{\mathrm{dist}}\approx3$ at $\tau\approx0.002$.
The fast decay of the static peak also indicates that particles are not moving simultaneously through
the fluid for a longer time period, but their motion is only positively
correlated during their collision.
In order to understand the dynamic peaks, imagine two particles undergoing
a collision at $\tau=0$. One particle (the self particle) is located at $r=0$
and has a positive velocity and the other particle is located at $r\approx1$.
The second particle would be pushed towards larger distances and move
towards a third particle located at $r\approx2$. The collision
of these particles at a later time $\tau>0$ would push back the second
particle at $r\approx1$ resulting in a negative velocity creating
a negative peak of $J_{rr}$ at $r\approx1$. The third particle
however would be pushed forward resulting in the observed positive
peak at $r\approx2$. With this mechanism the velocity of the self
particle is transferred across two particle diameters within a time
of $\tau\approx0.002$, at which the peak of the second correlation
shell attains its maximum. This would correspond to a propagation
speed of $v_{\mathrm{p}}\approx10^{3}$. We expect also weaker
peaks at outer correlation shells, which cannot be resolved with
the present data due to the noise. 
The estimate of the propagation speed is two orders of magnitude greater
than the velocity of the starting signal for the decay of the density correlations, 
$v_{\mathrm{int}}\approx10$.
This difference indicates that the ``starting signal'' is not transmitted
by a chain of collisions and supports the notion of transmission
via a slower mechanism, such as the diffusive dislocation of the self
particle and of the inner correlation shells as described above.

The transversal-transversal component of the static distinct current-current
correlator, $J_{tt}^{\mathrm{dist}}$, (shown in Fig.~\ref{fig:J2-dist}b)\hspace{0.03cm})
shows only a single negative static peak with a minimum value of $J_{tt}^{\mathrm{dist}}=-18$
at $r=1.02$. In contrast to the radial-radial component this
peak is negative and smaller by a factor of 20.
The peak being negative indicates shearing motion, in which particles tend to orbit each other rather than moving alongside of each other.
The peak decays within the same time scale as the peak of the radial-radial component, but we do not find dynamic peaks here.

\subsection{Superadiabatic van Hove current\label{sub:Results-Jad}}

\begin{figure}
\includegraphics[width=8.6cm,angle=0]{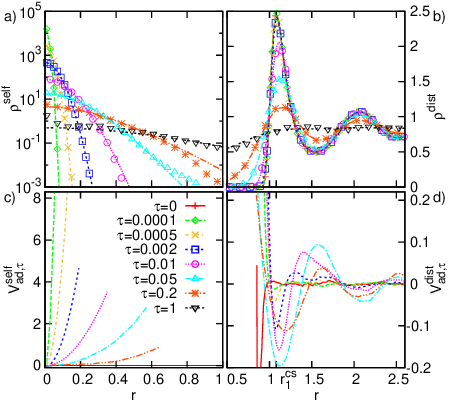}
\caption{\label{fig:Vad}a) and b) Comparison between the one-body densities in the adiabatic systems,
$\rho_{\mathrm{ad,\tau}}^{\alpha}(r)$, (symbols) and in the dynamic
test particle system, $\rho_{\mathrm{tp}}^{\alpha}(r,\tau)$, (lines)
as functions of distance $r$ at time differences from $\tau=0.0001$
to 1 as indicated; a) density of the self particle on logarithmic
scale and b) density of the distinct particles. $\rho_{\mathrm{ad,\tau}}^{\alpha}(r)$
is obtained from the adiabatic construction as described in Sec.\ \ref{sub:Superadiabatic-Force},
$\rho_{\mathrm{tp}}^{\alpha}(r,\tau)$ is obtained from the data of
the bulk equilibrium system via Eqs.\ (\ref{eq:rho-noneq-2}) and
(\ref{eq:rho-noneq-1-1}). $\rho_{\mathrm{ad,\tau}}^{\alpha}(r)$
is shown only for a selection of distances for clarity. 
Corresponding c) self and d) distinct external adiabatic potentials $V_{\mathrm{ad,\tau}}^{\mathrm{self}}$ and $V_{\mathrm{ad,\tau}}^{\mathrm{dist}}$
as functions of distance $r$ at time differences from $\tau=0.0001$
to 0.2 as indicated. $V_{\mathrm{ad,\tau}}^{\mathrm{\alpha}}$ is only shown
for values of $r$, at which we attained good agreement of $\rho_{\mathrm{ad,\tau}}^{\alpha}(r)$
and $\rho_{\mathrm{tp}}^{\alpha}(r,\tau)$.
$r_{1}^{\mathrm{cs}}$ in panels b) and d) indicates the position of the first maximum
of $\rho_{2}^{\mathrm{dist}}$.}
\end{figure}

In order to discriminate the dynamical effects that are driven by free energy changes from those that are purely dissipative,
we use the splitting into adiabatic and superadiabatic forces, as outlined in Sec.\ \ref{sub:Superadiabatic-Force}.
We adjusted the self and distinct adiabatic potentials $V_{\mathrm{ad,\tau}}^{\mathrm{self}}$
and $V_{\mathrm{ad,\tau}}^{\mathrm{dist}}$ to obtain an equilibrium one-body density,
which is equal to the one-body density in the dynamic TPL.
Figure \ref{fig:Vad}a) and b) show the comparisons of the self and distinct density distributions in the test particle system and the adiabatic system.
At times ranging from $\tau=0$ to 0.2 the iterations converged and the densities of both systems match very well.
An exception is the self particle density at large distances,
as seen on the logarithmic scale in Fig.\ \ref{fig:Vad}a). However,
the absolute density deviations in these regions are smaller than $10^{-2}$.
Hence, we assume that the deviations have only little effect on the
behavior of the entire system and conclude that the obtained adiabatic
potentials are valid in the regions where the densities converged.
For $\tau>0.2$ the iterations did not converge properly with
the iteration parameters used (as exemplarily shown for $\tau=1$
in Fig.\ \ref{fig:Vad}a) and b)\hspace{0.03cm}). We attribute this to the high degree of delocalization of the self particle.

The obtained self and distinct adiabatic potentials resulting from the adiabatic construction are shown in Fig.~\ref{fig:Vad}c) and d).
At $\tau=0$ the self adiabatic potential $V_{\mathrm{ad,\tau}}^{\mathrm{self}}(r)$ is an infinitely deep and infinitely narrow potential well
that fixes the self particle at the origin and thus generates the static TPL.
The distinct adiabatic potential $V_{\mathrm{ad,\tau}}^{\mathrm{dist}}(r)$ is zero in this case.
The non-zero values of $V_{\mathrm{ad,\tau}}^{\mathrm{dist}}(r)$ for $r<1$ (i.e.\ the distances, where the pair potential is strongly repulsive)
are an artifact of the spatial discretization of the adiabatic potential.
For $\tau>0$ the potential well of $V_{\mathrm{ad,\tau}}^{\mathrm{self}}(r)$
attains finite values and broadens and consequently generates the
broadening density distribution of the diffusing particle. $V_{\mathrm{ad,\tau}}^{\mathrm{dist}}(r)$
develops a repulsive core at small distances that reaches its maximum
strength at $\tau\approx0.002$ (see blue dashed line in Fig.~\ref{fig:Vad}d)\hspace{0.03cm}). For greater times the core starts
to vanish again as the distinct density approaches a uniform distribution.
An oscillatory tail develops that holds the correlation shells in place, although the self particle has already started
to move.

\begin{figure}
\includegraphics[width=8.6cm,angle=0]{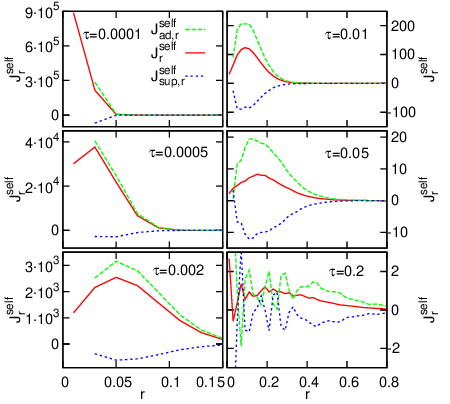}
\caption{\label{fig:Comparison-Jad-self}Comparison between the radial components of the self parts of
the total van Hove current, $J_{r}^{\mathrm{self}}$, of the adiabatic
van Hove current, $J_{\mathrm{ad,}r}^{\mathrm{self}}$, and of the
superadiabatic van Hove current, $J_{\mathrm{sup,}r}^{\mathrm{self}}$,
as functions of distance $r$ at time differences from $\tau=0.0001$
to 0.2 as indicated.}
\end{figure}

The radial components of the adiabatic and superadiabatic van Hove currents, $J_{\mathrm{ad,}r}^{\alpha}$ and $J_{\mathrm{sup,}r}^{\alpha}$,
have been calculated from $V_{\mathrm{ad,\tau}}^{\alpha}$ via Eqs.\ (\ref{eq:Jad-theory}) and (\ref{eq:Jsup-theory}).
Figure \ref{fig:Comparison-Jad-self} shows the comparison between the self parts of the total van Hove current,
$J_{r}^{\mathrm{self}}$ (cf.\ Sec.\ \ref{sub:Results-Self}), of the adiabatic van Hove current, $J_{\mathrm{ad,}r}^{\mathrm{self}}$,
and of the superadiabatic van Hove current, $J_{\mathrm{sup,}r}^{\mathrm{self}}$, for a sequence of times.
$J_{\mathrm{ad,}r}^{\mathrm{self}}$ and $J_{r}^{\mathrm{self}}$ show the same qualitative behavior for all times.
For $\tau<0.002$ also the numerical value is approximately the same, and as a consequence $J_{\mathrm{sup,}r}^{\mathrm{self}}$ is approximately zero.
For larger times $J_{\mathrm{ad,}r}^{\mathrm{self}}$ becomes bigger than $J_{r}^{\mathrm{self}}$
and $J_{\mathrm{sup,}r}^{\mathrm{self}}$ attains non-zero negative values.

\begin{figure}
\includegraphics[width=8.8cm,angle=0]{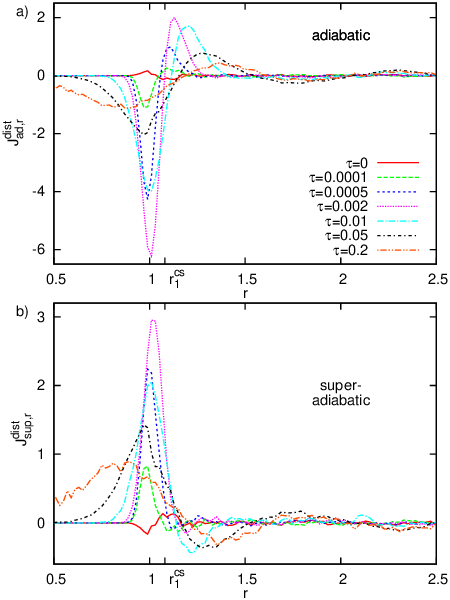}
\caption{\label{fig:Jad}Radial component of a) the adiabatic distinct van Hove current, $J_{\mathrm{ad,}r}^{\mathrm{dist}}$,
and b) of the superadiabatic distinct van Hove current, $J_{\mathrm{sup,}r}^{\mathrm{dist}}$, as functions of distance $r$ at time differences from $\tau=0$
to 0.2 as indicated. $r_{1}^{\mathrm{cs}}$
indicates the position of the first maximum of $\rho_{2}^{\mathrm{dist}}$.}
\end{figure}

The adiabatic and superadiabatic distinct van Hove currents are shown in Fig.\ \ref{fig:Jad}.
The adiabatic van Hove current $J_{\mathrm{ad,}r}^{\mathrm{dist}}$
(cf.\ Fig.\ \ref{fig:Jad}a)\hspace{0.03cm}) shows the same qualitative
behavior as the van Hove current $J_{r}^{\mathrm{dist}}$:
At $\tau=0$ the current is zero and for greater times there is a negative peak at $r\lesssim r_{1}^{\mathrm{cs}}$
and a positive peak at $r\gtrsim r_{1}^{\mathrm{cs}}$ (and also at outer correlation shells).
Again this behavior indicates a particle current from the correlation shells to the depletion zones in between.
However, the values of the adiabatic and the total van Hove currents are not equal and thus the superadiabatic contribution
$J_{\mathrm{sup,}r}^{\mathrm{dist}}$ (shown in Fig.~\ref{fig:Jad}b)\hspace{0.03cm}) is non-zero. 

\begin{figure}
\includegraphics[width=8.6cm,angle=0]{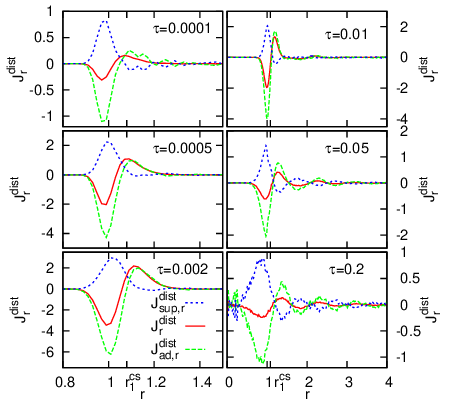}
\caption{\label{fig:Comparison-Jad}Comparison between the radial components of the distinct parts of
the total van Hove current, $J_{r}^{\mathrm{dist}}$, of the adiabatic
van Hove current, $J_{\mathrm{ad,}r}^{\mathrm{dist}}$, and of the
superadiabatic van Hove current, $J_{\mathrm{sup,}r}^{\mathrm{dist}}$,
as functions of distance $r$ at time differences from $\tau=0.0001$
to 0.2 as indicated. $r_{1}^{\mathrm{cs}}$ indicates the position
of the first maximum of $\rho_{2}^{\mathrm{dist}}$.}
\end{figure}

A detailed comparison between the distinct van Hove current and its adiabatic
and superadiabatic contributions is shown in Fig.\ \ref{fig:Comparison-Jad}.
The peaks of $J_{\mathrm{ad,}r}^{\mathrm{dist}}$ have the same signs
as the peaks of $J_{r}^{\mathrm{dist}}$ at the same distances. However,
the peaks of $J_{\mathrm{ad,}r}^{\mathrm{dist}}$ are larger than
the peaks of $J_{r}^{\mathrm{dist}}$. Hence, the peaks of $J_{\mathrm{sup,}r}^{\mathrm{dist}}$
have opposite signs to the peaks of $J_{\mathrm{ad,}r}^{\mathrm{dist}}$
and $J_{r}^{\mathrm{dist}}$. For $\tau<0.01$ this applies only
for the first negative peak at $r\lesssim r_{1}^{\mathrm{cs}}$. The
other peaks of $J_{\mathrm{ad,}r}^{\mathrm{dist}}$ start to exceed
the peaks of $J_{r}^{\mathrm{dist}}$ only for larger times. This
behavior reflects that the adiabatic dynamics overestimate the relaxation
speed of the density profile, which is a well known drawback of the
adiabatic DDFT, found in a variety of situations (see e.g.\ \citep{DDFT99,DDFT-HRods,Fortini-et-al}).
The deviations are often attributed to an approximate free energy
functional or discrepancies between canonical and grand canonical
dynamics. However, in our method the total van Hove current and the
adiabatic van Hove current were both calculated from canonical BD
simulations. Thus, both of these effects are absent. We conclude that the
adiabatic construction itself does not capture all features of the
dynamic behavior of the system. For an explanation of the discrepancy
between the adiabatic and the total van Hove current consider the
following picture. If a particle has moved in a certain direction,
it is more likely that there are other particles in front of that
particle than behind it. Figuratively speaking particles tend to have
a high density `bow wave' at front and a low density `wake' at back.
As the adiabatic construction has no memory of the dynamic trajectories, these kind of correlations are not taken into account.
We consider the absence of the repulsive effect of these bow wave particles
as one reason for the overestimated relaxation velocities in adiabatic theories. 

\section{Conclusion\label{sec:Conclusion}}

We have examined the dynamics of a LJ bulk fluid close to the triple point by means
of dynamic two-body correlation functions and identified the underlying decay mechanisms.
The comparison with the ideal gas showed that the qualitative behavior
of the self parts of the correlation functions can be rationalized in terms of diffusive motion alone.
The interactions between the particles inhibit the diffusion of particles and therefore
lead to stronger localized distributions, but do not change the functions qualitatively.
We find it quite remarkable that this applies not only for the two-body density but also for the van Hove current and for the current-current correlator.

The distinct correlation functions -- although linked via continuity equations
-- relax on different time scales and thus are affiliated to different processes.
The density correlations measured with the two-body density decay on the time scale of the Brownian time.
The plotting on a logscale revealed a novel decay mechanism, characterized by the seemingly paradoxical situation that
the dynamic correlation length is larger than the static correlation
length. The reason for this relation is the delayed decay of the
outer correlation shells. It remains an open task to develop a theoretical
framework for the dynamic decay of the correlations.
One possible route to this objective is employing the non-equilibrium Ornstein-Zernike equation \citep{NOZ}.
Another worthwhile project would be the comparisons of our findings with experiments of Statt $et\,al.$ \citep{Statt_Pinchaipat_16},
which provide trajectories of dispersed particles,
and with a modified version of DDFT presented by Stopper $et\,al.$ \citep{Stopper_Marolt_15,Stopper_Roth_15},
that yields correct long-time diffusion coefficients.

The velocity correlations measured with
the current-current correlator decay on the time scale of single collisions.
We were able to ascribe the different peaks of the current-current
correlator to different kinds of collisions occurring in the system.
Consequently, the current-current correlator is considered to be useful
to examine the microscopic motion of particles in many-body systems.
In the bulk system examined in this work, particularly information
about the transversal motion is not revealed solely from the two-body density
or the van Hove current. For collective phenomena and in the long-time limit,
BD and molecular dynamics are known to produce equivalent
results. However, this is not automatically implied for microscopic
motion. The current-current correlator appears to be a promising tool
to compare the two types of dynamics. Further investigations can also be
done by comparing our results for the current-current correlator to
the velocity autocorrelation function measured by van Megen and coworkers (see e.g.\ \citep{current-corr-exp,current-corr-simu}).

Furthermore, we split the van Hove current in its adiabatic and
superadiabatic contribution by employing the recently developed adiabatic construction for the test particle situation.
It is quite remarkable that there is a non-zero superadiabatic contribution to the equilibrium
bulk dynamics, that even exceeds the magnitude of the van Hove current.
This finding suggests that the validity of the adiabatic approximation is highly questionable in any high density fluid,
most likely also in inhomogeneous and non-equilibrium systems.
We want to reemphasize that all calculations in this work
are performed with canonical simulations. Hence, we conclude that
the quantitative differences between canonical BD and grand canonical
adiabatic DDFT do not only arise from ensemble differences but rather
from the adiabatic assumption in DDFT. In PFT the superadiabatic contribution
is taken into account via memory effects. Consequently, PFT can be
expected to produce improved results compared to DDFT. A possible
next step in this framework would be to obtain the memory functions
from the two-body correlation functions via the non-equilibrium Ornstein-Zernike
equations \citep{NOZ}. Furthermore
 it would be worthwhile to pursue the question whether the methods
 of obtaining the free power dissipation from functional line
 integration \citep{FLI} could be applied to the present
 problem. Non-Markovian effects were recently identified (in a
 system of one-dimensional hard rods) via memory kernels acting on
 the one-body current \citep{kernel}. Also the effects of ensemble differences
 of grand canonical versus canonical systems \citep{can-GC,PCD} could be
 relevant for the correct theoretical description of test particle
 situations.

\begin{acknowledgments}
We thank J.~Garhammer, A.~Fortini, D.~de las Heras, J.~M.~Brader, R.~Evans, and S.~Huth for useful discussions and comments. 
\end{acknowledgments}

\section{Appendix\label{sec:Appendix}}

\subsection{Divergences of vector and tensor fields in spherical coordinates\label{sub:Del*JvH}}

\subsubsection*{Divergence of a vector field}

A vector $\mathbf{v}$ in a spherical coordinate system, with local
unit vectors $\mathbf{e}_{r}$, $\mathbf{e}_{\theta}$, and $\mathbf{e}_{\varphi}$
defined in Sec.\ \ref{sub:Symm-and-Cosy},
can be written as
\begin{equation}
\mathbf{v}=\mathbf{e}_{r}v_{r}+\mathbf{e}_{\theta}v_{\theta}+\mathbf{e}_{\varphi}v_{\varphi}.\label{eq:vector-in-spher-cosy}
\end{equation}
The divergence operator $\nabla\cdot$ reads

\begin{equation}
\nabla\cdot=\frac{1}{r^{2}}\partial_{r}(r^{2}\mathbf{e}_{r}\cdot)+\frac{1}{r\sin\theta}\partial_{\theta}(\sin\theta\mathbf{e}_{\theta}\cdot)+\frac{1}{r\sin\theta}\partial_{\varphi}\mathbf{e}_{\varphi}\cdot.\label{eq:divergence-in-spher-cosy}
\end{equation}
Combining Eqs.\ (\ref{eq:vector-in-spher-cosy})
and (\ref{eq:divergence-in-spher-cosy}) we obtain for the divergence
of the vector field
\begin{equation}
\nabla\cdot\mathbf{v}=\frac{\partial_{r}(r^{2}v_{r})}{r^{2}}+\frac{\partial_{\theta}(\sin\theta v_{\theta})}{r\sin\theta}+\frac{\partial_{\varphi}v_{\varphi}}{r\sin\theta}.
\end{equation}
In case of a vector, that has only a non-zero radial component this
simplifies to
\begin{equation}
\nabla\cdot\mathbf{v}=\frac{\partial_{r}(r^{2}v_{r})}{r^{2}}.\label{eq:Del*v}
\end{equation}

\subsubsection*{Divergence of a tensor field}

A tensor $\underline{\underline{\mathbf{T}}}$ can be expressed as
a linear combination of dyadic products of unit vectors,
\begin{align}
\underline{\underline{\mathbf{T}}}= & \,\mathbf{e}_{r}\mathbf{e}_{r}T_{rr}+\mathbf{e}_{r}\mathbf{e}_{\varphi}T_{r\varphi}+\mathbf{e}_{r}\mathbf{e}_{\theta}T_{r\theta}\nonumber \\
 & +\mathbf{e}_{\varphi}\mathbf{e}_{r}T_{\varphi r}+\mathbf{e}_{\varphi}\mathbf{e}_{\varphi}T_{\varphi\varphi}+\mathbf{e}_{\varphi}\mathbf{e}_{\theta}T_{\varphi\theta}\\
 & +\mathbf{e}_{\theta}\mathbf{e}_{r}T_{\theta r}+\mathbf{e}_{\theta}\mathbf{e}_{\varphi}T_{\theta\varphi}+\mathbf{e}_{\theta}\mathbf{e}_{\theta}T_{\theta\theta}.\nonumber 
\end{align}
In case of a diagonal tensor, the equation simplifies to
\begin{equation}
\underline{\underline{\mathbf{T}}}=\mathbf{e}_{r}\mathbf{e}_{r}T_{rr}+\mathbf{e}_{\varphi}\mathbf{e}_{\varphi}T_{\varphi\varphi}+\mathbf{e}_{\theta}\mathbf{e}_{\theta}T_{\theta\theta}.
\end{equation}
Employing the divergence operator (\ref{eq:divergence-in-spher-cosy})
on $\underline{\underline{\mathbf{T}}}$, one first performs a scalar
product of the unit vectors of $\nabla$ with the first vector of
each dyadic product, yielding
\begin{align}
\nabla\cdot\underline{\underline{\mathbf{T}}}= & \,\frac{1}{r^{2}}\partial_{r}(r^{2}\mathbf{e}_{r}T_{rr})+\frac{1}{r\sin\theta}\partial_{\theta}(\sin\theta\mathbf{e}_{\theta}T_{\theta\theta})\nonumber\\
& +\frac{1}{r\sin\theta}\partial_{\varphi}(\mathbf{e}_{\varphi}T_{\varphi\varphi}).\label{eq:Del*T1}
\end{align}
While carrying out the partial derivatives, one should have in mind
that some of the partial derivatives of the unit vectors are non-zero.
The matrix of derivatives is given by
\begin{eqnarray}
\partial_{r}\mathbf{e}_{r}=0,\,\,\,\, & \partial_{\theta}\mathbf{e}_{r}=\mathbf{e}_{\theta},\,\, & \,\,\,\,\,\partial_{\varphi}\mathbf{e}_{r}=\mathbf{e}_{\varphi}\sin\theta,\nonumber \\
\partial_{r}\mathbf{e}_{\theta}=0,\,\,\,\, & \,\,\,\partial_{\theta}\mathbf{e}_{\theta}=-\mathbf{e}_{r}, & \,\,\,\,\,\partial_{\varphi}\mathbf{e}_{\theta}=\mathbf{e}_{\varphi}\cos\theta,\label{eq:d_ie_i}\\
\partial_{r}\mathbf{e}_{\varphi}=0,\,\,\,\, & \partial_{\theta}\mathbf{e}_{\varphi}=0,\,\,\,\,\, & \,\,\,\hspace{0.1cm}\partial_{\varphi}\mathbf{e}_{\varphi}=-\mathbf{e}_{r}\sin\theta-\mathbf{e}_{\theta}\cos\theta.\nonumber 
\end{eqnarray}
If furthermore the components of the tensor only depend on the radial
coordinate $r$ but not on the angular coordinates $\theta$ and $\varphi$,
then from Eqs.\ (\ref{eq:Del*T1}) and
(\ref{eq:d_ie_i}) we obtain
\begin{align}
\nabla\cdot\underline{\underline{\mathbf{T}}}=&\ \frac{\mathbf{e}_{r}}{r^{2}}\partial_{r}(r^{2}T_{rr})+\mathbf{e}_{\theta}\frac{\cos\theta}{r\sin\theta}T_{\theta\theta}\nonumber\\
&-\mathbf{e}_{r}\frac{T_{\theta\theta}}{r}-\mathbf{e}_{r}\frac{T_{\varphi\varphi}}{r}-\mathbf{e}_{\theta}\frac{\cos\theta}{r\sin\theta}T_{\varphi\varphi}\nonumber\\
=&\ \mathbf{e}_{r}\left[\frac{1}{r^{2}}\partial_{r}(r^{2}T_{rr})-\frac{T_{\theta\theta}}{r}-\frac{T_{\varphi\varphi}}{r}\right]\label{eq:del*T}\\
&+\mathbf{e}_{\theta}\frac{\cos\theta}{r\sin\theta}(T_{\theta\theta}-T_{\varphi\varphi}).\nonumber
\end{align}

\subsubsection*{Tensor divergence of a diagonal tensor}

We obtain the second divergence of a diagonal tensor $\nabla\nabla:\underline{\underline{\mathbf{T}}}$
by applying the divergence operator (\ref{eq:divergence-in-spher-cosy})
on the divergence of the tensor given in Eq.\ (\ref{eq:del*T}):
\begin{align}
\nabla\nabla:\underline{\underline{\mathbf{T}}}=&\ \nabla\cdot(\nabla\cdot\underline{\underline{\mathbf{T}}})\nonumber \\
=&\ \frac{1}{r^{2}}\partial_{r}(r^{2}(\frac{1}{r^{2}}\partial_{r}(r^{2}T_{rr})-\frac{T_{\theta\theta}}{r}-\frac{T_{\varphi\varphi}}{r}))\label{eq:DelDel**T}\\
&+\frac{1}{r\sin\theta}\partial_{\theta}(\sin\theta\frac{\cos\theta}{r\sin\theta}(T_{\theta\theta}-T_{\varphi\varphi}))\nonumber \\
=&\ \frac{1}{r^{2}}\left[\partial_{r}^{2}(r^{2}T_{rr})-\partial_{r}(rT_{\theta\theta}+rT_{\varphi\varphi})-(T_{\theta\theta}-T_{\varphi\varphi})\right]\nonumber
\end{align}
Again we assume the components of the tensor to only depend on $r$.

\subsection{Continuity equations\label{sub:CEq-numerical}}

\begin{figure}[t]
\hspace{-0.2cm}
\includegraphics[width=8.7cm,angle=0]{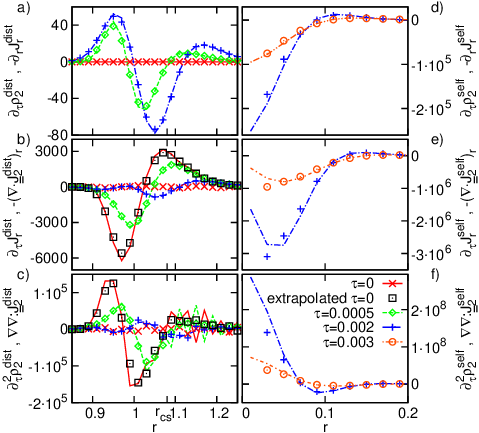}
\caption{\label{fig:CEq}Numerical data of the temporal derivatives and divergences
of the two-body correlation functions as functions of distance $r$
as appearing in the two-body continuity equations. Panels a) -- c) show
distinct parts of functions at time differences $\tau=0$, 0.0005,
and 0.002, panels d) -- f) show self parts at time differences $\tau=0.002$
and 0.003. a) and d) show the temporal derivative of $\rho_{2}^{\alpha}$
and the negative divergence of $\mathbf{J}_{\text{\ensuremath{\mathrm{vH}}}}^{\alpha}$
(cf.\ Eq.\ (\ref{eq:ceq1-1})\hspace{0.03cm}). b) and e) show the
radial components of the temporal derivative of $\mathbf{J}_{\text{\ensuremath{\mathrm{vH}}}}^{\alpha}$
and of the negative divergence of $\JJ^{\alpha}$
(cf.\ Eq.\ (\ref{eq:ceq3-1})\hspace{0.03cm}). c) and f) show the
second temporal derivative of $\rho_{2}^{\alpha}$ and the second
divergence of $\JJ^{\alpha}$ (cf.\ Eq.\ (\ref{eq:ceq5-1})\hspace{0.03cm}).
Lines show temporal derivatives, symbols show divergences. The black
squares in panels b) and c) show the first and second divergences
of $\JJ^{\mathrm{dist}}(\mathbf{r},0)$,
as extrapolated from non-zero time arguments. $r_{1}^{\mathrm{cs}}$
indicates the position of the first maximum of $\rho_{2}^{\mathrm{dist}}$.}
\end{figure}

We verified the self consistency of the obtained correlation functions
by numerically calculating their derivatives from our simulation data
as appearing in the continuity equations derived in Sec.\ \ref{sub:CEq-theory}.
As mentioned before, the divergences of the spherically symmetrical
functions can be calculated from the spherical components as derived
in the previous section. The appearing first and second partial derivatives
of a particular function $f(s)$ are computed via symmetrical differentiation
according to
\begin{equation}
\partial_{s}f(s)=\frac{f(s+\Delta s)-f(s-\Delta s)}{2\Delta s},
\end{equation}
\begin{equation}
\partial_{s}^{2}f(s)=\frac{f(s+\Delta s)-2f(s)+f(s-\Delta s)}{\Delta s^{2}},
\end{equation}
where $\Delta s$ is the width of the spatial or temporal grid of
the numerical data. For $\tau>0$ we find good agreement of the numerical
data with the continuity equations (\ref{eq:ceq1-1}), (\ref{eq:ceq3-1}),
and (\ref{eq:ceq5-1}) as shown in Fig.\ \ref{fig:CEq}.
Differences of the derivatives of the self parts (Figs.\ \ref{fig:CEq}d) --
f)\hspace{0.03cm}) at low distances arise from the breakdown of
the numerical differentiation near the singularity of the spherical
coordinate system. For $\tau=0$ we find the distinct current-current
correlator $\JJ^{\mathrm{dist}}$
to be discontinuous. This is not consistent with the distinct two-body
density $\rho_{2}^{\mathrm{dist}}$ and the van Hove current $\mathbf{J}_{\mathrm{vH}}^{\mathrm{dist}}$
being continuous at $\tau=0$, as can be seen in Figs.\ \ref{fig:CEq}b)
and c). The discrepancy is assumed to be an artifact due to the discretization
of time in the BD simulation. In order to retain the consistency of the correlation
functions we extrapolate the current-current correlator at $\tau=0$
from non-zero time arguments. The divergences of the extrapolated values
are also shown in Figs.\ \ref{fig:CEq}b) and c) and show good agreement
with the related temporal derivatives of $\mathbf{J}_{\mathrm{vH}}^{\mathrm{dist}}$
and $\rho_{2}^{\mathrm{dist}}$. The presented results for $\JJ^{\mathrm{dist}}(\mathbf{r},0)$
in Sec.\ \ref{sec:Results} are the extrapolated values.

\subsection{Self two-body correlation functions for non-interacting particles\label{sub:J-self-re}}

\begin{figure}
\includegraphics[width=8.8cm,angle=0]{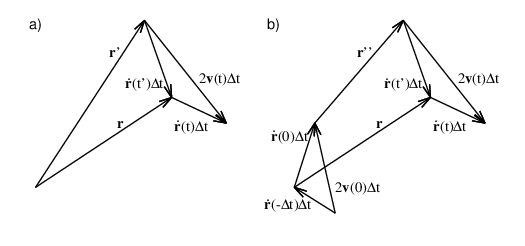}
\vspace{10 mm}
\caption{\label{fig:sketch-J-re}Sketch of the movement of a particle in discrete time steps. Symbols
are defined in the text.}
\end{figure}

In order to obtain analytical solutions for the self van Hove current $\mathbf{J}_{\mathrm{vH}}^{\mathrm{self}}(\mathbf{r},t)$
and the self current-current correlator $\JJ^{\mathrm{self}}(\mathbf{r},t)$
for freely diffusing particles at non-zero time arguments, we consider
movement in discrete time steps $\Delta t$ after Eq.\ (\ref{eq:trajectory-1})
and perform a limit $\Delta t\rightarrow0$ to obtain the exact solution.
We assume that a particle starts at the origin $\mathbf{r}=0$ at
time $t=0$ without loss of generality. The particle reaches position
$\mathbf{r}'\equiv\mathbf{r}-\dot{\mathbf{r}}(t')\Delta t$ at time
$t'\equiv t-\Delta t$ before it finally reaches position $\mathbf{r}$
at time $t$ with displacement velocity $\dot{\mathbf{r}}(t')$ as
shown in Fig.\ \ref{fig:sketch-J-re}a).
From there the particle moves with displacement velocity $\dot{\mathbf{r}}(t)$.

The probability to reach $\mathbf{r}'$ at time $t'$ is given by
the two-body density $\rho_{2}^{\mathrm{self}}(\mathbf{r}',t')$.
The probability distribution of random displacement velocities is
(according to Eqs.\ (\ref{eq:ran_Force-1})
and (\ref{eq:Var})\hspace{0.03cm})
\begin{equation}
p(\dot{\mathbf{r}})=(2\pi\sigma_{v}^{2})^{-3/2}\exp\left(-\frac{\dot{\mathbf{r}}^{2}}{2\sigma_{v}^{2}}\right).\label{eq:velocity-distribution}
\end{equation}
The (forward) van Hove current can now be calculated as an integral
over the displacement velocities $\dot{\mathbf{r}}(t')$ and $\dot{\mathbf{r}}(t)$
(weighted with their probability distributions) of the symmetrical
velocity $\mathbf{v}(t)$ multiplied with the two-body density $\rho_{2}^{\mathrm{self}}(\mathbf{r}',t')$:

\begin{align}
\mathbf{J}_{\mathrm{vH}}^{\mathrm{self}}(\mathbf{r},t)=\lim_{\Delta t\rightarrow0} & \intg\mathrm{d}\dot{\mathbf{r}}(t')p(\dot{\mathbf{r}}(t'))\label{eq:intg-JvH-self-re}\\
&\times\intg\mathrm{d}\dot{\mathbf{r}}(t)p(\dot{\mathbf{r}}(t))\mathbf{v}(t)\rho_{2}^{\mathrm{self}}(\mathbf{r}',t').\nonumber
\end{align}
In the limit of small time steps $\Delta t$ we can make a first order
Taylor expansion of the two-body density around $\mathbf{r}$ and
$t$:
\begin{equation}
\rho_{2}^{\mathrm{self}}(\mathbf{r}',t')\approx\rho_{2}^{\mathrm{self}}(\mathbf{r},t)-\dot{\mathbf{r}}(t')\Delta t\cdot\nabla\rho_{2}^{\mathrm{self}}(\mathbf{r},t)-\Delta t\partial_{t}\rho_{2}^{\mathrm{self}}(\mathbf{r},t).
\end{equation}
The symmetrical velocity is related to the displacement
velocities via Eq.\ (\ref{eq:v-rdot}), the two-body density $\rho_{2}^{\mathrm{self}}(\mathbf{r},t)$
is given by Eq.\ (\ref{eq:rho2-self-re}). Hence, the integral in
(\ref{eq:intg-JvH-self-re}) can be calculated analytically and one
obtains\begin{equation}
\mathbf{J}_{\mathrm{vH}}^{\mathrm{self}}(\mathbf{r},t)=\frac{\mathbf{r}}{2t}\rho_{2}^{\mathrm{self}}(r,t).\label{eq:J-re-apend}
\end{equation}

The current-current correlator can also be calculated with this method
as an average of $\mathbf{v}(0)\mathbf{v}(t)\rho_{2}^{\mathrm{self}}(\mathbf{r}'',t'')$
with $\mathbf{r}''\equiv\mathbf{r}-[\dot{\mathbf{r}}(0)+\dot{\mathbf{r}}(t')]\Delta t$
and $t''\equiv t-2\Delta t$ (cf.\ Fig.\ \ref{fig:sketch-J-re}b)\hspace{0.03cm}).
Therefore, integrations over $\dot{\mathbf{r}}(-\Delta t)$, $\dot{\mathbf{r}}(0)$,
$\dot{\mathbf{r}}(t')$, and $\dot{\mathbf{r}}(t)$ have to be carried
out yielding
\begin{align}
\JJ^{\mathrm{self}}(\mathbf{r},t)=\lim_{\Delta t\rightarrow0}&\intg\mathrm{d}\dot{\mathbf{r}}(-\Delta t)p(\dot{\mathbf{r}}(-\Delta t))\intg\mathrm{d}\dot{\mathbf{r}}(0)p(\dot{\mathbf{r}}(0))\label{eq:J2-self-re-intg}\nonumber\\
&\times\intg\mathrm{d}\dot{\mathbf{r}}(t')p(\dot{\mathbf{r}}(t'))\intg\mathrm{d}\dot{\mathbf{r}}(t)p(\dot{\mathbf{r}}(t))\\
&\times\mathbf{v}(0)\mathbf{v}(t)\rho_{2}^{\mathrm{self}}(\mathbf{r}'',t'').\nonumber
\end{align}
The first order terms of the Taylor expansion vanish while carrying
out the integrals. Hence, a second order Taylor expansion of the two-body
density is needed, which reads
\begin{align}
\rho_{2}^{\mathrm{self}}(\mathbf{r}'',t'')\approx&\ \rho_{2}^{\mathrm{self}}(\mathbf{r},t)-\left[\dot{\mathbf{r}}(0)+\dot{\mathbf{r}}(t')\right]\Delta t\cdot\nabla\rho_{2}^{\mathrm{self}}(\mathbf{r},t)\nonumber \\
&-2\Delta t\partial_{t}\rho_{2}^{\mathrm{self}}(\mathbf{r},t)\nonumber\\
&+\frac{1}{2}\left\{ \left[\dot{\mathbf{r}}(0)+\dot{\mathbf{r}}(t')\right]\Delta t\cdot\nabla\right\} ^{2}\rho_{2}^{\mathrm{self}}(\mathbf{r},t)\label{eq:2nd-order-Taylor}\\
&+2\Delta t\partial_{t}\left[\dot{\mathbf{r}}(0)+\dot{\mathbf{r}}(t')\right]\Delta t\cdot\nabla\rho_{2}^{\mathrm{self}}(\mathbf{r},t)\nonumber\\
&+\frac{1}{2}\left(2\Delta t\partial_{t}\right)^{2}\rho_{2}^{\mathrm{self}}(\mathbf{r},t).\nonumber 
\end{align}
Substituting Eqs.\ (\ref{eq:velocity-distribution}) and (\ref{eq:2nd-order-Taylor})
into Eq.\ (\ref{eq:J2-self-re-intg}) and carrying out the integrals,
the current-current correlator reads
\begin{equation}
\JJ^{\mathrm{self}}(\mathbf{r},t)=\frac{1}{4t^{2}}\left(\mathbf{r}\mathbf{r}-\sigma_{t}^{2}\underline{\underline{\mathbf{I}}}\right)\rho_{2}^{\mathrm{self}}(r,t).\label{eq:J2-re-apend}
\end{equation}
Equations (\ref{eq:J-re-apend}) and (\ref{eq:J2-re-apend}) are the
analytic solutions for the van Hove current and the current-current
correlator at non-zero time arguments as used for the freely relaxing
reference system in Eqs.\ (\ref{eq:JvH-self-re}) and (\ref{eq:J2-self-re}).

\bibliography{Bibtex/Bibtex}

\begin{thebibliography}{10}

\bibitem{DFT-Mermin}
N.~D.\ Mermin,
\href{https://doi.org/10.1103/PhysRev.137.A1441}{Phys.\ Rev.\ {\bf 137}, A1441 (1965)}.

\bibitem{DFT-Evans79}
R.\ Evans,
\href{https://doi.org/10.1080/00018737900101365}{Adv.\ Phys.\ {\bf 28}, 143 (1979)}.

\bibitem{DFT_overview_16}
R.~Evans, M.~Oettel, R.~Roth, and G.~Kahl,
\href{https://doi.org/10.1088/0953-8984/28/24/240401}{J.\ Phys.:\ Condens.\ Matter {\bf 28}, 240401 (2016)}.

\bibitem{DDFT99}
U.~M.~B.\ Marconi and P.~Tarazona,
\href{https://doi.org/10.1063/1.478705}{J.\ Chem.\ Phys.\ {\bf 110}, 8032 (1999)}.

\bibitem{Archer-Evans04}
A.~J.\ Archer and R.\ Evans,
\href{https://doi.org/10.1063/1.1778374}{J.\ Chem.\ Phys.\ {\bf 121}, 4246 (2004)}.

\bibitem{DDFT-HRods}
F.~Penna and P.~Tarazona,
\href{https://doi.org/10.1063/1.2189243}{J.\ Chem.\ Phys.\ {\bf 124}, 164903 (2006)}.

\bibitem{Fortini-et-al}
A.~Fortini, D.~de~las Heras, J.~M.\ Brader, and M.~Schmidt,
\href{https://doi.org/10.1103/PhysRevLett.113.167801}{Phys.\ Rev.\ Lett.\ {\bf 113}, 167801 (2014)}.

\bibitem{dTPL}
A.~J.\ Archer, P.~Hopkins, and M.~Schmidt,
\href{https://doi.org/10.1103/PhysRevE.75.040501}{Phys.\ Rev.\ E {\bf 75}, 040501(R) (2007)}.

\bibitem{dTPL2}
P.~Hopkins, A.~Fortini, A.~J.\ Archer, and M.~Schmidt,
\href{https://doi.org/10.1063/1.3511719}{J.\ Chem.\ Phys.\ {\bf 133}, 224505 (2010)}.

\bibitem{GvH}
L.~van Hove,
\href{https://doi.org/10.1103/PhysRev.95.249}{Phys.\ Rev.\ {\bf 95}, 249 (1954)}.

\bibitem{Stopper_Marolt_15}
D.~Stopper, K.~Marolt, R.~Roth, and H.~Hansen-Goos,
\href{https://doi.org/10.1103/PhysRevE.92.022151}{Phys.\ Rev.\ E {\bf 92}, 022151 (2015)}.

\bibitem{Stopper_Roth_15}
D.~Stopper, R.~Roth, and H.~Hansen-Goos,
\href{https://doi.org/10.1063/1.4935967}{J.\ Chem.\ Phys.\ {\bf 143}, 181105 (2015)}.

\bibitem{PFT13}
M.~Schmidt and J.~M.\ Brader,
\href{https://doi.org/10.1063/1.4807586}{J.\ Chem.\ Phys.\ {\bf 138}, 214101 (2013)}.

\bibitem{NOZ}
J.~M. Brader and M.~Schmidt,
\href{https://doi.org/10.1063/1.4820399}{J. Chem. Phys. {\bf 139}, 104108 (2013)}.

\bibitem{active}
P.~Krinninger, M.~Schmidt, and J.~M.~Brader,
\href{https://doi.org/10.1103/PhysRevLett.117.208003}{Phys.\ Rev.\ Lett.\ {\bf 117}, 208003 (2017)}.

\bibitem{PFT-DTPL15}
J.~M.\ Brader and M.~Schmidt,
\href{https://doi.org/10.1088/0953-8984/27/19/194106}{J.\ Phys.:\ Condens.\ Matter {\bf 27}, 194106 (2015)}.

\bibitem{PFT14}
J.~M.\ Brader and M.~Schmidt,
\href{https://doi.org/10.1063/1.4861041}{J.\ Chem.\ Phys.\ {\bf 140}, 034104 (2014)}.

\bibitem{LJ-PD-rc4.4}
A.~Trokhymchuk and J.~Alejandre,
\href{https://doi.org/10.1063/1.480192}{J.\ Chem.\ Phys.\ {\bf 111}, 8510 (1999)}.

\bibitem{LJ}
J.~K.\ Johnson, J.~A.\ Zollweg, and K.~E.\ Gubbins,
\href{https://doi.org/10.1080/00268979300100411}{Mol.\ Phys.\ {\bf 78}, 591 (1993)}.

\bibitem{hansen_mcdonald}
J.-P.\ Hansen and I.~R.\ McDonald,
\href{https://doi.org/10.1016/C2010-0-66723-X}{\it Theory of Simple Liquids} (Academic Press, London, 4th edition, 2013).

\bibitem{video-microscopy-Valentine}
M.~T.\ Valentine, P.~D.\ Kaplan, D.~Thota, J.~C.\ Crocker, T.~Gisler, R.~K.\ Prud'homme, M.~Beck, and D.~A.\ Weitz,
\href{https://doi.org/10.1103/PhysRev.64.061506}{Phys.\ Rev.\ {\bf 64}, 061506 (2001)}.

\bibitem{video-microscopy-Murray}
C.~A.\ Murray and D.~G.\ Grier,
\href{https://doi.org/10.1146/annurev.physchem.47.1.421}{Annu.\ Rev.\ Phys.\ Chem.\ {\bf 47}, 421 (1996)}.

\bibitem{transient}
P.~Krinninger, A.~Fortini, and M.~Schmidt,
\href{https://doi.org/10.1103/PhysRevE.93.042601}{Phys.\ Rev.\ E {\bf 93}, 042601 (2016)}.

\bibitem{decay93}
R.~Evans, J.~R.\ Henderson, D.~C.\ Hoyle, A.~O.\ Parry, and Z.~A.\ Sabeur,
\href{https://doi.org/10.1080/00268979300102621}{Mol.\ Phys.\ {\bf 80}, 755 (1993)}.

\bibitem{Asymptotic1}
R.~Evans, R.~J.~F.\ Leote~de Carvalho, and J.~R.\ Henderson,
\href{https://doi.org/10.1063/1.466920}{J.\ Chem.\ Phys.\ {\bf 100}, 591 (1994)}.

\bibitem{FW}
M.~E.\ Fisher and B.~Widom,
\href{https://doi.org/10.1063/1.1671624}{J.\ Chem.\ Phys.\ {\bf 50}, 3756 (1969)}.

\bibitem{Statt_Pinchaipat_16} 
A.~Statt, R.~Pinchaipat, F.~Turci, R.~Evans, and C.~P.~Royall,
\href{https://doi.org/10.1063/1.4945808}{J.\ Chem.\ Phys {\bf 144}, 144506 (2016)}.

\bibitem{Klapp_Zeng_08}
S.~H.~L.~Klapp, Y.~Zeng, D.~Qu, and R.~von Klitzing,
\href{https://doi.org/10.1103/PhysRevLett.100.118303}{Phys.\ Rev.\ Lett.\ {\bf 100}, 118303 (2008)}.

\bibitem{current-corr-exp}
V.~A.\ Martinez, E.~Zaccarelli, E.~Sanz, C.~Valeriani, and W.~van Megen,
\href{https://doi.org/10.1038/ncomms6503}{Nat.\ Commun.\ {\bf 5}, 5503 (2014)}.

\bibitem{frencel_smit}
D.~Frenkel and B.~Smit,
\href{https://doi.org/10.1016/B978-0-12-267351-1.X5000-7}
{\it Understanding Molecular Simulation}{\it\ - From Algorithms to Applications}
	(Academic Press, London, 2nd edition, 2002).

\bibitem{Gauss}
P.~J.\ Rossky, J.~D.\ Doll, and H.~L.\ Friedman,
\href{https://doi.org/10.1063/1.436415}{J.\ Chem.\ Phys.\ {\bf 69}, 4628 (1978)}.

\bibitem{LJ-TrP}
E.~A.\ Mastny and J.~J.\ de~Pablo,
\href{https://doi.org/10.1063/1.2753149}{J.\ Chem.\ Phys.\ {\bf 127}, 104504 (2007)}.

\bibitem{Bernreuther_Schmidt}
E.~Bernreuther and M.~Schmidt,
\href{https://doi.org/10.1103/PhysRevE.94.022105}{Phys.\ Rev.\ E 94, 022105 (2016)}.

\bibitem{STPL}
J.~K.\ Percus,
\href{https://doi.org/10.1103/PhysRevLett.8.462}{Phys.\ Rev.\ Lett.\ {\bf 8}, 462 (1962)}.

\bibitem{ES}
A.~Einstein,
\href{https://doi.org/10.1002/andp.200590009}{Ann.\ Physik {\bf 19}, 371 (1906)}.

\bibitem{Henderson-Sabeur92}
J.~R.\ Henderson and Z.~A.\ Sabeur,
\href{https://doi.org/10.1063/1.463652}{J.\ Chem.\ Phys.\ {\bf 97}, 6750 (1992)}.

\bibitem{FW-HSAY}
W.~E.\ Brown, R.~J.~F.\ Leote~de Carvalho, and R.~Evans,
\href{https://doi.org/10.1080/00268979650026541}{Mol.\ Phys.\ {\bf 88}, 579 (1996)}.

\bibitem{Dijkstra-Evans00}
M.~Dijkstra and R.~Evans,
\href{https://doi.org/10.1063/1.480598}{J.\ Chem.\ Phys.\ {\bf 112}, 1449 (2000)}.

\bibitem{Asymptotic2}
R.~J.~F.\ Leote~de Carvalho, R.~Evans, and Y.~Rosenfeld,
\href{https://doi.org/10.1103/PhysRevE.59.1435}{Phys.\ Rev.\ E {\bf 59}, 1435 (1999)}.

\bibitem{Leote-de-Carvalho-Evans94}
R.~J.~F.\ Leote~de Carvalho, R.~Evans, D.~C.~Hoyle, and J.~R.~Henderson,
\href{https://doi.org/10.1088/0953-8984/6/44/008}{J.\ Phys.:\ Condens.\ Matter {\bf 6}, 9275 (1994)}.

\bibitem{current-corr-simu}
W.~van Megen, V.~A.\ Martinez, and G.~Bryant,
\href{https://doi.org/10.1103/PhysRevLett.103.258302}{Phys.\ Rev.\ Lett.\ {\bf 103}, 258302 (2009)}.

\bibitem{FLI}
J.~M.~Brader, M.~Schmidt,
\href{https://doi.org/10.1080/00268976.2015.1042086}{Mol.\ Phys.\ {\bf 113}, 2873 (2015)}.

\bibitem{kernel}
A.~Fortini, D.~de~las Heras, J.~M.\ Brader, and M.~Schmidt
(to be published).

\bibitem{can-GC}
D.~de las Heras, M.~Schmidt,
\href{https://doi.org/10.1103/PhysRevLett.113.238304}{Phys.\ Rev.\ Lett.\ {\bf 113}, 238304 (2014)}.

\bibitem{PCD}
D.~de las Heras, J.~M.~Brader, A.~Fortini, M.~Schmidt,
\href{https://doi.org/10.1088/0953-8984/28/24/244024}{J.\ Phys.:\ Condens.\ Matter {\bf 28}, 244024 (2016)}.

\end{thebibliography}

\end{document}